\documentclass{jfm}
\usepackage{amssymb}
\usepackage{amsfonts}
\usepackage{amsmath}
\usepackage{euscript}
\usepackage{frcursive}
\usepackage{color}
\usepackage{graphicx}
\usepackage{rotating}
\usepackage{bm}
\usepackage{curves}
\usepackage[svgnames,dvipsnames]{pstricks}
\usepackage{lettrine}
\usepackage{mathrsfs}
\usepackage{booktabs}
\usepackage{scalefnt}
\usepackage[normalem]{ulem}
\usepackage{dcolumn}
\usepackage{natbib}
\usepackage{tikz}
\DeclareGraphicsExtensions{{},.pdf}
\providecommand{\tsn}[1]{{\text{\scalefont{0.80}#1}}}

\providecommand{\tabref}[1]{{\textup{(Tab.~\ref{#1})}}}
\providecommand{\figref}[1]{{\textup{(Fig.~\ref{#1})}}}

\providecommand{\figrefnp}[1]{{\textup{Fig.~\ref{#1}}}}

\providecommand{\eqrefsab}   [2]{\textup{(\ref{#1}, \ref{#2})}}

\providecommand{\tabrefsab}   [2]{{\textup{(Tabs.~\ref{#1}, \ref{#2})}}}
\providecommand{\figrefsab}   [2]{{\textup{(Figs.~\ref{#1}, \ref{#2})}}}

\providecommand{\figrefsatob} [2]{{\textup{(Figs.~\ref{#1}--\ref{#2})}}}
\providecommand{\parref}[1]{{\textup{(\S\ref{#1})}}}

\providecommand{\parrefnp}[1]{{\textup{\S\ref{#1}}}}

\providecommand{\const}{{\rm const}}

\providecommand{\ie}{{\em ie}}
\providecommand{\eg}{{\em eg}}
\providecommand{\viz}{{\em viz}}
\providecommand{\vs}{{\em vs}}

\newcommand{\abs}[1]{\left\lvert#1\right\rvert}

\providecommand{\TKE}{{\rm k}}
\newcommand{\tsr}[1]{{\boldsymbol{\mathbf{#1}}}}

\newcommand{\tr}{\mathrm{tr}}
\newcommand{\I}[1]{  \mathrm{I}_{\boldsymbol{\mathbf{#1}}}}
\newcommand{\II}[1]{ \mathrm{II}_{\boldsymbol{\mathbf{#1}}}}
\newcommand{\III}[1]{\mathrm{III}_{\boldsymbol{\mathbf{#1}}}}

\makeatletter
\newcommand{\vast}{\bBigg@{4}}
\newcommand{\Vast}{\bBigg@{5}}
\makeatother
\def\NumERICCS{N\kern-.09em\lower.5ex\hbox{\tsn{um}}\kern-.09em\tsn{ERICCS}}
\title[Dissipation tensor $\varepsilon_{ij}$]{The dissipation tensor $\varepsilon_{ij}$ in wall turbulence}
\author[G.A. Gerolymos and I. Vallet]
{G.\ns A.\ns G\ls E\ls R\ls O\ls L\ls Y\ls M\ls O\ls S,\ns
I.\ns V\ls A\ls L\ls L\ls E\ls T\footnote{Email address for correspondence: isabelle.vallet@upmc.fr}}
\affiliation{\scalefont{0.94}{Sorbonne Universit\'es, Universit\'e Pierre-et-Marie-Curie (UPMC), 4 place Jussieu, 75005 Paris, France}}
\date{\today}

\begin{document}

\maketitle

\begin{abstract}
The paper investigates the dissipation tensor $\varepsilon_{ij}$ in wall turbulence. Available \tsn{DNS} data are examined to illustrate the differences in the anisotropy of the dissipation tensor $\varepsilon_{ij}$ with respect to the anisotropy of the 
Reynolds-stresses $r_{ij}$. The budgets of the transport equations of the dissipation tensor $\varepsilon_{ij}$ are studied using novel \tsn{DNS} data of low-Reynolds-number turbulent plane channel flow with spatial resolution sufficiently fine to accurately
determine the correlations of products of 2-derivatives of fluctuating velocities $u_i'$ and pressure $p'$ which appear in various terms.
Finally, the influence of the Reynolds number on the diagonal components of $\varepsilon_{ij}$ ($\varepsilon_{xx}$, $\varepsilon_{yy}$, $\varepsilon_{zz}$) and on the various terms in their transport equations is studied using available \tsn{DNS} data of
Vreman and Kuerten ({\it Phys. Fluids} {\bf 26} (2014) 085103).
\end{abstract}


%
%
%
%
%
%
%
%
%
\section{Introduction}\label{DTWT_s_I}
%
%
%
%
%
%
%
%
%

The Reynolds-stress tensor $-\rho r_{ij}$ (where $r_{ij}:=\overline{u_i'u_j'}$ are the 2-moments of the fluctuating velocity components, $u_i\in\{u,v,w\}$ are the velocity components in a Cartesian reference-frame $x_i\in\{x,y,z\}$, $\overline{(\cdot)}$
denotes Reynolds (ensemble) averaging and $(\cdot)'$ turbulent fluctuations around the mean-value) represents the influence of turbulent momentum-mixing on the mean-flow \citep[pp. 86--87]{Pope_2000a}.
Throughout the paper we consider incompressible flow (with density $\rho\approxeq\const$), in an inertial reference-frame \citep{Speziale_1989a},
with a Newtonian constitutive relation for the viscous stresses \citep[(2.4), p. 31]{Davidson_2004a}, and we assume that the variations of dynamic viscosity are negligible ($\mu\approxeq\const$).
The exact equation for the transport of $r_{ij}$ \citep[(7.178--7.181), p. 315]{Pope_2000a}
\begin{alignat}{6}
\underbrace{\rho\dfrac{\partial\overline{u'_i u_j'}}
                      {\partial                  t }+\rho\bar u_\ell\dfrac{\partial\overline{u'_i u_j'}}
                                                                          {\partial              x_\ell}}_{\displaystyle C_{ij}}=&\underbrace{\dfrac{\partial       }
                                                                                                                                                    {\partial x_\ell}\Bigg(\mu\dfrac{\partial\overline{u'_iu_j'}}
                                                                                                                                                                                    {\partial x_\ell            }\Bigg)}_{\displaystyle d_{ij}^{(\mu)}}
                                                                                                                                + \underbrace{\dfrac{\partial       }
                                                                                                                                                    {\partial x_\ell}\Big(-\rho\overline{u'_iu_j'u'_\ell}\Big)}_{\displaystyle d_{ij}^{(u)}}
                                                                                                                                + \underbrace{\Bigg(-\overline{u_i'\dfrac{\partial p' }
                                                                                                                                                                         {\partial x_j}}
                                                                                                                                                    -\overline{u_j'\dfrac{\partial p' }
                                                                                                                                                                         {\partial x_i}}\Bigg)}_{\displaystyle \Pi_{ij}}
                                                                                                                                    \notag\\
                                                                                                                                +&\underbrace{\Bigg(-\rho\overline{u_i'u_\ell'}\dfrac{\partial\bar u_j}
                                                                                                                                                                                     {\partial  x_\ell}
                                                                                                                                                    -\rho\overline{u_j'u_\ell'}\dfrac{\partial\bar u_i}
                                                                                                                                                                                     {\partial  x_\ell}\Bigg)}_{\displaystyle P_{ij}}
                                                                                                                                - \underbrace{\Bigg(2\mu\overline{\dfrac{\partial u_i'  }
                                                                                                                                                                        {\partial x_\ell}
                                                                                                                                                                  \dfrac{\partial u_j'  }
                                                                                                                                                                        {\partial x_\ell}}\Bigg)}_{\displaystyle\rho\varepsilon_{ij}}
                                                                                                                                    \label{Eq_DTWT_s_I_001}
\end{alignat}
states that the convection of $r_{ij}$ by the mean-velocity field $C_{ij}$ is the balance of 5 mechanisms:
diffusion by molecular viscosity $d_{ij}^{(\mu)}$,
diffusion by the fluctuating-velocity field $d_{ij}^{(u)}$,
interaction of the fluctuating velocity with the fluctuating pressure-gradient $\Pi_{ij}$,
production by the interaction of the Reynolds-stresses with mean velocity-gradients $P_{ij}$,
and destruction by molecular friction $-\rho\varepsilon_{ij}$.
Notice that deterministic body-forces (\eg\ gravity in incompressible flow) do not appear in \eqref{Eq_DTWT_s_I_001} (\ie\ do not influence directly $r_{ij}$).
Budgets of \eqref{Eq_DTWT_s_I_001} in wall-turbulence have been studied via \tsn{DNS} \citep{Kim_2012a} by various authors,
for plane channel \citep*{Mansour_Kim_Moin_1988a,
                          Moser_Kim_Mansour_1999a},
pipe \citep*{ElKhoury_Schlatter_Noorani_Fischer_Brethouwer_Johansson_2013a}
or boundary-layer \citep*{Sillero_Jimenez_Moser_2013a} flows. On the other hand, although the dissipation tensor
\begin{subequations}
                                                                                                                                    \label{Eq_DTWT_s_I_002}
\begin{alignat}{6}
\varepsilon_{ij}:=2\nu\overline{\dfrac{\partial u_i'  }
                                      {\partial x_\ell}
                                \dfrac{\partial u_j'  }
                                      {\partial x_\ell}}
                                                                                                                                    \label{Eq_DTWT_s_I_002a}
\end{alignat}
has been examined \citep{Mansour_Kim_Moin_1988a}
as part of the budgets of \eqref{Eq_DTWT_s_I_001}, and with respect to its anisotropy \citep*{Lai_So_1990a,
                                                                                              Jovanovic_Ye_Durst_1995a},
to the authors knowledge, the transport equations for $\varepsilon_{ij}$ in wall turbulence have not been studied in detail.
Instead, attention has concentrated on its half-trace
\begin{alignat}{6}
\varepsilon:=\tfrac{1}{2}\varepsilon_{mm}\stackrel{\eqref{Eq_DTWT_s_I_002a}}{=}\nu\overline{\dfrac{\partial u_m'  }
                                                                                                  {\partial x_\ell}
                                                                                            \dfrac{\partial u_m'  }
                                                                                                  {\partial x_\ell}}>0
                                                                                                                                    \label{Eq_DTWT_s_I_002b}
\end{alignat}
\end{subequations}
which represents the dissipation-rate of the turbulent kinetic energy $\TKE:=\tfrac{1}{2}r_{mm}$ and is strictly positive being the average of a sum of squares of real numbers \citep{Schumann_1977a}.
In \eqref{Eq_DTWT_s_I_002} $\nu:=\rho^{-1}\;\mu$ is the kinematic viscosity.
Notice that although most authors \citep{Mansour_Kim_Moin_1988a,
                                         Lai_So_1990a,
                                         Speziale_Gatski_1997a,
                                         Moser_Kim_Mansour_1999a,
                                         Pope_2000a,
                                         Jakirlic_Hanjalic_2002a,
                                         Hoyas_Jimenez_2008a}
use definition \eqref{Eq_DTWT_s_I_002a} for $\varepsilon_{ij}$,
there are some workers in the field \citep{Jovanovic_Ye_Durst_1995a,
                                           Oberlack_1997a} who do not include the factor $2$ in \eqref{Eq_DTWT_s_I_002a},
with corresponding absence of the factor $\tfrac{1}{2}$ in \eqref{Eq_DTWT_s_I_002b}.

In an important early work, \citet{Lee_Reynolds_1987a} analysed their \tsn{DNS} computations of homogeneous turbulence
(distorted by different types of strain and at different nondimensional mean strain-rates $2\mathrm{k}\varepsilon^{-1}\sqrt{\tfrac{1}{2}\overline{S}_{ij}\overline{S}_{ij}}$ and then left unstrained to relax)
using \tsn{AIM} (anisotropy invariant mapping) of the 2-moment tensors of the fluctuating fields of 
velocity $r_{ij}:=\overline{u_i'u_j'}$, dissipation $\varepsilon_{ij}$ \eqref{Eq_DTWT_s_I_002b} and vorticity \citep[$\vec{\omega}:=\mathrm{rot}{\vec{V}}$, pp. 39-50]{Davidson_2004a} $\overline{\omega_i'\omega_j'}$, with the underlying idea that 
$\overline{u_i'u_j'}$ characterizes the anisotropy of the large structures whereas $\varepsilon_{ij}$ and $\overline{\omega_i'\omega_j'}$ measure the anisotropy of the small scales. They concluded that in the distorsion phase 
"{\em the vorticity field $\overline{\omega_i'\omega_j'}$ is always more anisotropic than the velocity field $\overline{u_i'u_j'}$}" \citep[p. 60]{Lee_Reynolds_1987a} and that in the relaxation-phase 
"{\em the large-scale anisotropy is also coupled with the small-scale turbulence}" \citep[p. 66]{Lee_Reynolds_1987a}.
\citet{Durbin_Speziale_1991a} also considered $\varepsilon_{ij}$ \eqref{Eq_DTWT_s_I_002b} as the representative "{\em small-scale statistic}" and disproved (in an informal but plausible analysis by contradiction) the idea of local small-scale isotropy
when the large scales are subjected to high nondimensional mean strain-rates, independently of the Reynolds number.
\citet{Oberlack_1997a} reviews different experimental and computational results concluding that "{\em \tsn{DNS} have revealed the strong nonisotropic nature of the dissipation process}",
and that "{\em a finite level of small-scale anisotropy must always exist if the large scales are anisotropic}".
In such situations of highly anisotropic $\varepsilon_{ij}$ "{\em the small scales are not dominated by a classical energy cascade}" \citep[p. 9]{Bhattacharya_Kassinos_Moser_2008a}.
The importance of the small-scale anisotropy represented by the anisotropy of $\varepsilon_{ij}$ \eqref{Eq_DTWT_s_I_002b} is central in the modelling work of \citet{Speziale_Gatski_1997a}
and \citet{Lumley_Yang_Shih_1999a}. In an early work, \citet*{Tagawa_Nagano_Tsuji_1991a} developed a 12-equation $r_{ij}$--$\varepsilon_{ij}$ closure to impove near-wall modelling.

Wall turbulence is much more complex, not only because of the strong inhomogeneity in the wall-normal direction which is induced by the mean-flow field \citep{Buschmann_GadelHak_2007a},
but also because of the direct influence of the wall, which is twofold: on the one hand the no-slip boundary-condition imposes, at the wall, a 2-C \citep[2-component]{Lumley_1978a,
                                                                                                                                                                       Simonsen_Krogstad_2005a}
componentality \citep{Kassinos_Reynolds_Rogers_2001a} both on the 2-moments of the fluctuating velocities $r_{ij}:=\overline{u_i'u_j'}$ \citep[Fig. 17, p. 32]{Mansour_Kim_Moin_1988a}
and on the dissipation tensor $\varepsilon_{ij}$ \citep[Fig. 18, p. 32]{Mansour_Kim_Moin_1988a},
and on the other hand wall-echo \citep*{Kim_1989a,
                                        Chang_Piomelli_Blake_1999a,
                                        Gerolymos_Senechal_Vallet_2013a}
strongly impacts the fluctuating-pressure field.

\citet{Mansour_Kim_Moin_1988a} presented for the first time, in incompressible fully developed (streamwise invariant in the mean) plane channel flow,  the budgets of the transport equation for $\varepsilon$ \eqref{Eq_DTWT_s_I_002b},
which had "{\em eluded measurement techniques}".
The analysis of \citet{Mansour_Kim_Moin_1988a} reveals the specific behaviour of the 4 different production mechanisms of $\varepsilon$ \citep[(23), p. 23]{Mansour_Kim_Moin_1988a},
and \tsn{DNS} data show \citep[Fig. 6, p. 24]{Mansour_Kim_Moin_1988a} that all of these mechanisms are of comparable importance near the wall (inner-scaled wall-distance $y^+\lessapprox 15$), except for the production by the
mean-velocity Hessian $P^{(3)}_\varepsilon:=-2\rho\nu\overline{u_k'\partial_{x_j}u_i'}\partial^2_{x_kx_j}\bar u_i$ \citep[(23), p. 23]{Mansour_Kim_Moin_1988a} which is generally weaker.
On the other hand, further away from the wall ($y^+\gtrapprox30$) the production by the triple correlations of the fluctuating velocity-gradients
$P^{(4)}_\varepsilon:=-2\rho\nu\overline{\partial_{x_k}u_i'\partial_{x_j}u_i'\partial_{x_j}u_k'}$ \citep[(23), p. 23]{Mansour_Kim_Moin_1988a}
becomes the dominant $\varepsilon$-production mechanism. The predominance of $P^{(4)}_\varepsilon$ away from the wall is consistent with the quasi-homogeneous order-of-magnitude analysis of
\citet[pp. 88--92]{Tennekes_Lumley_1972a} which concludes that the major production mechanism of the fluctuating vorticity $\overline{\omega_i'\omega_i'}$
is $\rho\overline{\omega_i'\omega_j'S_{ij}'}$ (where $S_{ij}:=\tfrac{1}{2}(\partial_{x_j}u_i+\partial_{x_i}u_j)$ is the strain-rate tensor) in line with \citet{Taylor_1938a}.
Near the wall, the main hypothesis of this order-of-magnitude analysis \citep[pp. 88--92]{Tennekes_Lumley_1972a}, \viz\ that the lengthscale characterizing the mean velocity-gradients is much larger
than some appropriately defined microscale \citep{Taylor_1938a,
                                                  Kolovandin_Vatutin_1972a,
                                                  Vreman_Kuerten_2014a}
characterizing the fluctuating velocity-gradients, obviously breaks down \citep[Fig. 7, p. 15]{Vreman_Kuerten_2014a}:
"{\em near walls $\cdots$ the scales of energy containing motions and the scales of dissipative motions are the same}" \citep[p. 510]{Rodi_Mansour_1993a}.

The wall-asymptotic (as $y^+\to0$) behaviour of various turbulent correlations can be studied by Taylor-series expansions \citep[\S4.6, pp. 136--141]{Riley_Hobson_Bence_2006a}
in the wall-normal direction $y^+$, in inner scaling \citep[$\cdot^+$]{Buschmann_GadelHak_2007a},
of the fluctuating velocities $u_i^+$ and fluctuating pressure $p'^+$, under the constraints of the no-slip condition and of the Navier-Stokes equations.
This procedure is described in \citet[pp. 620--621]{Hinze_1975a},
and appears in a less systematic form (related to the mean-velocity expansion) in \citet[p. 163]{Townsend_1976a}
and \citet[p. 271]{Monin_Yaglom_1971a}.
\citet{Chapman_Kuhn_1986a} used this approach to resolve a controversy that existed at that time \citep{Patel_Rodi_Scheuerer_1985a}
concerning the asymptotic behaviour $-\overline{u'^+v'^+}\propto{y^+}^3$ of the turbulent shear stress near the wall ($y^+\to0$).
\citet{Launder_Reynolds_1983a} applied this technique to determine the wall-asymptotic behaviour of the components of the dissipation tensor $\varepsilon_{ij}$.
\citet{Mansour_Kim_Moin_1988a} studied the wall-asymptotics
of the various terms in the $r_{ij}$-transport budgets and in the budgets of the transport equation \citep[(23), p. 23]{Mansour_Kim_Moin_1988a} for the half-trace $\varepsilon$ \eqref{Eq_DTWT_s_I_002b}.
\citet[Tab. 3, p. 191]{Hanjalic_1994a} reported the wall-asymptotic expansions of the $r_{ij}$-anisotropy tensor $2b_{ij}$ \eqref{Eq_DTWT_s_DNSDepsij_ss_A_001a}.

\citet{Jovanovic_Ye_Durst_1995a} also focus on the transport equation for the half-trace $\varepsilon$ \eqref{Eq_DTWT_s_I_002b},
arguing that the components of the dissipation tensor $\varepsilon_{ij}$ can be "{\em analytically interpreted in terms of its trace $\varepsilon_{\ell\ell}$ and of the 2-point velocity correlations}".
Using standard inhomogeneous 2-point-correlation techniques \citep[pp. 43--44]{Chou_1945a},
$\varepsilon_{ij}$ \eqref{Eq_DTWT_s_I_002a} is split into an inhomogeneous part
(which is $\tfrac{1}{2}\nu\nabla^2r_{ij}$ and vanishes at the limit of homogeneous turbulence) and a quasi-homogeneous part which is always present.
\citet{Jakirlic_Hanjalic_2002a} modelled the unclosed terms in the corresponding equation for the homogeneous part of $\varepsilon$ \eqref{Eq_DTWT_s_I_002b}, \viz\ $\varepsilon-\tfrac{1}{2}\nu\nabla^2\mathrm{k}$,
which they used in a Reynolds-stress 7-equation model framework, further developed and applied by \citet{Jakirlic_Eisfeld_JesterZurker_Kroll_2007a} and \citet{Jakirlic_Maduta_2015a}.

Recently, \citet{Vreman_Kuerten_2014b} have studied using \tsn{DNS} the statistics, including skewness and flatness \citep[(3.37), p. 43]{Pope_2000a} and pdfs \citep[probability-density functions, pp. 39--41]{Pope_2000a}
of the variance of the components of the fluctuating velocity-gradient $\overline{(\partial_{x_j}u_i')^2}$ and Hessian $\overline{(\partial^2_{x_jx_k}u_i')^2}$, and of analogous correlations for the fluctuating
pressure ($\overline{(\partial_{x_i}p')^2}$, $\overline{(\partial^2_{xx}p')^2}$, $\overline{(\partial^2_{yy}p')^2}$, $\overline{(\partial^2_{zz}p')^2}$), in plane channel flow at friction Reynolds number \eqref{Eq_DTWT_s_AppendixABVSy+0_ss_WUs_001g}
$Re_{\tau_w}\approxeq590$. They also analyzed the budgets of the transport equations for the 9 components $\overline{(\partial_{x_j}u_i')^2}$, which can be combined to obtain the transport equations for the diagonal components
of $\varepsilon_{ij}$ \eqref{Eq_DTWT_s_I_002a}, $\{\varepsilon_{xx},\varepsilon_{yy},\varepsilon_{zz}\}$. Correlations for the transport-equation budgets of the shear component $\varepsilon_{xy}$  
cannot be extracted from the data of \citet{Vreman_Kuerten_2014b}. Near the wall, the shear component $\varepsilon_{xy}$ is comparable with the wall-normal component $\varepsilon_{yy}$ and should therefore be taken into account when
studying the anisotropy of $\varepsilon_{ij}$ \eqref{Eq_DTWT_s_I_002a}, and it has been argued \citep*{Hanjalic_Jakirlic_1993a,
                                                                                                       Gerolymos_Lo_Vallet_Younis_2012a}
that correct modelling of the shear component $\varepsilon_{xy}$ is important in advanced \tsn{RANS} models.
Furthermore, the results of the present work indicate that the $\varepsilon_{xy}$-budgets behave unlike those of the diagonal components, and such specific behaviour
also applies to the relative importance of various mechanisms of production. An important observation of \citet{Vreman_Kuerten_2014b} is that the Laplacian$\nabla^2p'$ is highly intermittent near the wall \citep[Fig. 12, p. 21]{Vreman_Kuerten_2014b}
and this implies significant trailing tails in the corresponding pdfs \citep[Fig. 11, p. 20]{Vreman_Kuerten_2014b} and consequently very high values of flatness. Therefore correlations in the $\varepsilon_{ij}$-transport equations
containing the components of the fluctuating pressure-Hessian $\partial_{x_jx_k}^2 p'$ require long observation times to reach statistical convergence.
\citet{Vreman_Kuerten_2016a} produced similar data in a highly resolved \tsn{DNS} at $Re_{\tau_w}\approxeq180$.

The purpose of the paper is to study in detail the near-wall behaviour of the dissipation tensor $\varepsilon_{ij}$, using \tsn{DNS} data for incompressible low-$Re_{\tau_w}$
(friction Reynolds number) fully developed turbulent plane channel flow. The solver used to acquire these data was described in \citet*{Gerolymos_Senechal_Vallet_2010a} and uses an $O(\Delta\ell^{17})$ upwind-biased
discretization for the convective terms \citep*{Gerolymos_Senechal_Vallet_2009a}. It has been validated by systematic \citep{Gerolymos_Senechal_Vallet_2010a,
                                                                                                                             Gerolymos_Senechal_Vallet_2013a,
                                                                                                                             Gerolymos_Vallet_2014a}
comparison with standard \citep{Moser_Kim_Mansour_1999a,
                                delAlamo_Jimenez_2003a,
                                Hoyas_Jimenez_2006a,
                                Hoyas_Jimenez_2008a}
\tsn{DNS} data (profiles, 2-point correlations and spectra in the homogeneous $xz$-directions, and $r_{ij}$-transport budgets), including assessment of the influence of compressibility at the
low-Mach-number limit \citep[Appendix A, pp. 45--51]{Gerolymos_Senechal_Vallet_2013a}.
It should be stated from the outset that the present \tsn{DNS} data were obtained using a compressible airflow solver \citep{Gerolymos_Senechal_Vallet_2010a}, and computations were run at centerline Mach number
$\bar M_\tsn{CL}=0.35$. Specific comparisons with incompressible \tsn{DNS} data of \citet{Vreman_Kuerten_2014a,
                                                                                          Vreman_Kuerten_2014b,
                                                                                          Vreman_Kuerten_2016a}
presented in \parrefnp{DTWT_s_RetauwIoepsijBs} of the paper clearly demonstrate that the high-order statistics examined in the present work are not influenced by compressibility, at this $\bar M_\tsn{CL}=0.35$.

In \parrefnp{DTWT_s_DNSDepsij} we review existing \tsn{DNS} data concerning the anisotropy of $r_{ij}$ (large scales) and $\varepsilon_{ij}$ (small scales),
in fully developed plane channel flow, including wall-asymptotics and the influence of Reynolds number,
highlighting the main componentality differences between these 2 tensors.
In \parrefnp{DTWT_s_epsijBs} we present, to our knowledge for the first time,
the budgets of the transport equations (obtained by straightforward manipulations of the fluctuating continuity and momentum equations; \parrefnp{DTWT_s_epsijBs_ss_epsijTEq})
for all the components of $\varepsilon_{ij}$ (including the shear component $\varepsilon_{xy}$),
in low Reynolds number ($Re_{\tau_w}\approxeq180$) plane channel flow.
In \parrefnp{DTWT_s_RetauwIoepsijBs} we use the data of \citet{Vreman_Kuerten_2014a,
                                                               Vreman_Kuerten_2014b,
                                                               Vreman_Kuerten_2016a},
which can be combined to provide the budgets of the diagonal components of $\varepsilon_{ij}$ (but not those of the shear component), both to assess the $Re_{\tau_w}\approxeq180$ data and to discuss the evolution with $Re_{\tau_w}$ of different terms.
Finally, in \parrefnp{DTWT_s_C} we summarize the main new results obtained in the paper and discuss perspectives for future research.

%
%
%
%
%
%
%
%
%
\section{DNS data for $\varepsilon_{ij}$}\label{DTWT_s_DNSDepsij}
%
%
%
%
%
%
%
%
%

We consider fully developed turbulent plane channel flow, and use nondimensional inner variables (wall-units; \parrefnp{DTWT_s_AppendixABVSy+0_ss_WUs}).
The channel height is $2\delta$, $y$ is the wall-normal direction, $x$ and $z$ are, respectively, the streamwise and spanwise (crossflow) directions,
along which mean-flow velocities and all turbulent moments are invariant.
Available \tsn{DNS} databases \citep{Kim_Moin_Moser_1987a,
                                     Moser_Kim_Mansour_1999a,
                                     Hu_Morfey_Sandham_2002a,
                                     Hu_Morfey_Sandham_2003a,
                                     Hu_Morfey_Sandham_2006a,
                                     delAlamo_Jimenez_2003a,
                                     Hoyas_Jimenez_2006a,
                                     Hoyas_Jimenez_2008a,
                                     LozanoDuran_Jimenez_2014a,
                                     Bernardini_Pirozzoli_Orlandi_2014a,
                                     Vreman_Kuerten_2014a,
                                     Vreman_Kuerten_2014b,
                                     Vreman_Kuerten_2016a,
                                     Lee_Moser_2015a}
of turbulent plane channel flow provide information on the budgets of the $r_{ij}$-transport equations \eqref{Eq_DTWT_s_I_001}
which can be used to assess analogies and differences in the anisotropy of $r_{ij}$ and $\varepsilon_{ij}$ as the flow Reynolds number $Re_{\tau_w}$ \eqref{Eq_DTWT_s_AppendixABVSy+0_ss_WUs_001g} varies.

%
%
%
%
%
\subsection{Anisotropy}\label{DTWT_s_DNSDepsij_ss_A}
%
%
%
%
%

The traceless anisotropy tensor $b_{ij}$ \citep{Lumley_1978a} and its invariants \citep{Rivlin_1955a} $\II{b}$ and $\III{b}$ \citep{Gerolymos_Lo_Vallet_2012a}
\begin{subequations}
                                                                                                                                    \label{Eq_DTWT_s_DNSDepsij_ss_A_001}
\begin{align}
r_{ij}:=\overline{u_i'u_j'}\quad;\quad
\TKE:=\tfrac{1}{2}\overline{u_\ell'u_\ell'}\quad;\quad
b_{ij}:=\frac{\overline{u_i'u_j'}}{2\TKE}-\tfrac{1}{3}\delta_{ij}
                                                                                                                                    \label{Eq_DTWT_s_DNSDepsij_ss_A_001a}\\
\II{b}=-\tfrac{1}{2}b_{mk}b_{km}\quad;\quad
\III{b}=\tfrac{1}{3}b_{mk}b_{k\ell}b_{\ell m}\quad;\quad
A:=1+27\III{b}+9\II{b}
                                                                                                                                    \label{Eq_DTWT_s_DNSDepsij_ss_A_001b}
\end{align}
\end{subequations}
describe the local state of the Reynolds-stress tensor whose locus in the $(\III{b},-\II{b})$-plane lies within \citeauthor{Lumley_1978a}'s (\citeyear{Lumley_1978a}) realisability triangle,
determined by the fact \citep{Schumann_1977a} that the diagonal components ($r_{xx}:=\overline{u'^2}$, $r_{yy}:=\overline{v'^2}$, and $r_{zz}:=\overline{w'^2}$) of the Reynolds-stress tensor $\tsr{r}$ are positive in any reference-frame ($\tsr{r}$ is positive-definite).
The simplified form of $\II{b}$ and $\III{b}$ in \eqref{Eq_DTWT_s_DNSDepsij_ss_A_001} takes into account that $\tsr{b}$ is traceless ($\I{b}=\tr\tsr{b}\stackrel{\eqref{Eq_DTWT_s_DNSDepsij_ss_A_001}}{=}0$).
In \eqref{Eq_DTWT_s_DNSDepsij_ss_A_001}
$A\in[0,1]$ is \citeauthor{Lumley_1978a}'s (\citeyear{Lumley_1978a}) flatness parameter which \citep{Lumley_1978a} tends to zero at the 2-component limit \citep[\tsn{TCL}]{Craft_Launder_2001a},
\ie\ in the wall-turbulence case \citep{Launder_Shima_1989a} at the wall (inversely $A\to0$ implies \tsn{TCL}).

The mathematical proof \citep[pp. 138--140]{Lumley_1978a} that the invariants $\{\III{b},-\II{b}\}$ \eqref{Eq_DTWT_s_DNSDepsij_ss_A_001b} must lie within the realizability triangle only uses the condition that the diagonal components
of the symmetric Reynolds-stress tensor $\tsr{r}$ ($r_{ij}:=\overline{u_i'u_j'}$) are nonnegative for any orientation of the axes-of-coordinates \citep{Schumann_1977a} and hence also for the frame of the principal
axes where the symmetric tensor $\tsr{r}$ is diagonalized \citep[pp. 25--28]{Aris_1962a} with diagonal components its real eigenvalues. Since the eigenvalues of $\tsr{r}$ are real and nonnegative,
$\tsr{r}$ is positive-definite \citep[Theorem 2.3, p. 186]{Stewart_1998a}. Therefore, the realizability triangle applies to the anisotropy invariants of any
positive-definite symmetric order-2 tensor (equivalently any symmetric order-2 tensor whose diagonal elements are nonegative for every orientation of the axes-of-coordinates).
From definition \eqref{Eq_DTWT_s_I_002a}, the dissipation tensor $\tsr{\varepsilon}$ is also positive-definite because its diagonal components
($\varepsilon_{xx}:=2\nu\overline{\partial_{x_\ell}u'\partial_{x_\ell}u'}$, $\varepsilon_{yy}:=2\nu\overline{\partial_{x_\ell}v'\partial_{x_\ell}v'}$, and $\varepsilon_{zz}:=2\nu\overline{\partial_{x_\ell}w'\partial_{x_\ell}w'}$) are always positive.
Therefore, the corresponding traceless anisotropy tensor $b_{\varepsilon_{ij}}$ and its invariants \citep{Rivlin_1955a} $\II{b_\varepsilon}$ and $\III{b_\varepsilon}$
\begin{subequations}
                                                                                                                                    \label{Eq_DTWT_s_DNSDepsij_ss_A_002}
\begin{align}
\varepsilon_{ij}\stackrel{\eqref{Eq_DTWT_s_I_002a}}{:=}2\nu\overline{\dfrac{\partial u_i'  }
                                                                           {\partial x_\ell}
                                                                     \dfrac{\partial u_j'  }
                                                                           {\partial x_\ell}}\quad;\quad
\varepsilon\stackrel{\eqref{Eq_DTWT_s_I_002b}}{:=}\tfrac{1}{2}\varepsilon_{mm}\quad;\quad
b_{\varepsilon_{ij}}:=\frac{\varepsilon_{ij}}{2\varepsilon}-\tfrac{1}{3}\delta_{ij}
                                                                                                                                    \label{Eq_DTWT_s_DNSDepsij_ss_A_002a}\\
\II{b_\varepsilon}=-\tfrac{1}{2}b_{\varepsilon_{mk}}b_{\varepsilon_{km}}\quad;\quad
\III{b_\varepsilon}=\tfrac{1}{3}b_{\varepsilon_{mk}}b_{\varepsilon_{k\ell}}b_{\varepsilon_{\ell m}}\quad;\quad
A_\varepsilon:=1+27\III{b_\varepsilon}+9\II{b_\varepsilon}
                                                                                                                                    \label{Eq_DTWT_s_DNSDepsij_ss_A_002b}
\end{align}
\end{subequations}
define the same realizability triangle as $\tsr{b}$ \citep{Lumley_1978a,
                                                           Simonsen_Krogstad_2005a}.
This was first recognized by \citet[p. 56]{Lee_Reynolds_1987a}, who also state regarding the invariants \eqrefsab{Eq_DTWT_s_DNSDepsij_ss_A_001b}
                                                                                                                 {Eq_DTWT_s_DNSDepsij_ss_A_002b}
that "{\em each \tsn{AIM} \emph{(anisotropy invariant mapping)} has the boundaries first defined by \citet{Lumley_Newman_1977a} for the Reynolds-stress \tsn{AIM}}".

Since isotropy of $r_{ij}$ ($\varepsilon_{ij}$) corresponds to $\tsr{b}=0$ ($\tsr{b_\varepsilon}=0$), the larger the distance of the components $b_{ij}$ ($b_{\varepsilon_{ij}}$) from 0 the higher the anisotropy \citep{Simonsen_Krogstad_2005a}.
\begin{sidewaystable}
\vspace{6.0in}
\begin{center}
\rule{\textwidth}{.25pt}
\scalefont{.7}{
\begin{alignat}{8}
b_{xx}&\sim& \dfrac{2\overline{{A_u'^+}^2}-\overline{{A_w'^+}^2}}
                   {3\left(\overline{{A_u'^+}^2}+\overline{{A_w'^+}^2}\right)}
      &   +&2\dfrac{\overline{{A_w'^+}^2}\,\overline{A_u'^+B_u'^+}-\overline{{A_u'^+}^2}\,\overline{A_w'^+B_w'^+}}
                   {\left(\overline{{A_u'^+}^2}+\overline{{A_w'^+}^2}\right)^2}\,y^+
      &   +&   O({y^+}^2)
                                                                                                                                    \notag\\
b_{xy}&\sim&
      &    &\dfrac{\overline{A_u'^+B_v'^+}}
                  {\overline{{A_u'^+}^2}+\overline{{A_w'^+}^2}}\,y^+
      &   +&O({y^+}^2)
                                                                                                                                    \notag\\
b_{yy}&\sim&-\tfrac{1}{3}
      &    &
      &   +&\dfrac{\overline{{B_v'^+}^2}}
                  {\overline{{A_u'^+}^2}+\overline{{A_w'^+}^2}}\,{y^+}^2
      &   +&   O({y^+}^3)
                                                                                                                                    \notag\\
b_{yz}&\sim&
      &    &{\color{blue}{\vast[\dfrac{\overline{A_w'^+B_v'^+}}
                                {\overline{{A_u'^+}^2}+\overline{{A_w'^+}^2}}\,y^+}}
      &{\color{blue}{+}}
           &{\color{blue}{O({y^+}^2)\vast]}}
                                                                                                                                    \notag\\
b_{zz}&\sim& \dfrac{2\overline{{A_w'^+}^2}-\overline{{A_u'^+}^2}}
                   {3\left(\overline{{A_u'^+}^2}+\overline{{A_w'^+}^2}\right)}
      &    &-2\dfrac{\overline{{A_w'^+}^2}\,\overline{A_u'^+B_u'^+}-\overline{{A_u'^+}^2}\,\overline{A_w'^+B_w'^+}}
                    {\left(\overline{{A_u'^+}^2}+\overline{{A_w'^+}^2}\right)^2}\,y^+
      &   +& O({y^+}^2)
                                                                                                                                    \notag\\
b_{zx}&\sim&{\color{blue}{\Bigg[\dfrac{\overline{A_u'^+A_w'^+}}
                   {\overline{{A_u'^+}^2}+\overline{{A_w'^+}^2}}}}
      &   +&{\color{blue}{\dfrac{\left(\overline{{A_u'^+}^2}+\overline{{A_w'^+}^2}\right)\left(\overline{A_u'^+B_w'^+}+\overline{A_w'^+B_u'^+}\right)
                                -2\overline{A_u'^+A_w'^+}\left(\overline{A_u'^+B_u'^+}+\overline{A_w'^+B_w'^+}\right)}
                                {\left(\overline{{A_u'^+}^2}+\overline{{A_w'^+}^2}\right)^2}\,y^+}}
      &   +&{\color{blue}{ O({y^+}^2)\Bigg]}}
                                                                                                                                    \notag
\end{alignat}
\begin{alignat}{8}
\II{b}\sim&-\dfrac{\overline{{A_u'^+}^2}^2-\overline{{A_u'^+}^2}\,\overline{{A_w'^+}^2}+\overline{{A_w'^+}^2}^2{\color{blue}{+\left[3\overline{A_u'^+A_w'^+}^2\right]}}}{3\left(\overline{{A_u'^+}^2}+\overline{{A_w'^+}^2}\right)^2}
&+&\Vast(
2\dfrac{\left(\overline{{A_u'^+}^2}^2-\overline{{A_u'^+}^2}\,\overline{{A_w'^+}^2}\right)\overline{A_w'^+B_w'^+}+\left(\overline{{A_w'^+}^2}^2-\overline{{A_u'^+}^2}\,\overline{{A_w'^+}^2}\right)\overline{A_u'^+B_u'^+}}
       {\left(\overline{{A_u'^+}^2}+\overline{{A_w'^+}^2}\right)^3}
                                                                                                                                    \notag\\
&&
&{\color{blue}{+\left[\dfrac{4\overline{A_u'^+A_w'^+}^2\left(\overline{A_u'^+B_u'^+}+\overline{A_w'^+B_w'^+}\right)-2\overline{A_u'^+A_w'^+}\left(\overline{{A_u'^+}^2}+\overline{{A_w'^+}^2}\right)\left(\overline{A_u'^+B_w'^+}+\overline{A_w'^+B_u'^+}\right)}
              {\left(\overline{{A_u'^+}^2}+\overline{{A_w'^+}^2}\right)^3}\right]}}\Vast)y^++O({y^+}^2)
                                                                                                                                    \notag
\end{alignat}
\begin{alignat}{8}
\III{b}\sim&\dfrac{2\left(\overline{{A_u'^+}^2}^2+\overline{{A_w'^+}^2}^2\right)-5\overline{{A_u'^+}^2}\,\overline{{A_w'^+}^2}{\color{blue}{+\left[9\overline{A_u'^+A_w'^+}^2\right]}}}{27\left(\overline{{A_u'^+}^2}+\overline{{A_w'^+}^2}\right)^2}
&-&\Vast(
2\dfrac{\left(\overline{{A_u'^+}^2}^2-\overline{{A_u'^+}^2}\,\overline{{A_w'^+}^2}\right)\overline{A_w'^+B_w'^+}+\left(\overline{{A_w'^+}^2}^2-\overline{{A_u'^+}^2}\,\overline{{A_w'^+}^2}\right)\overline{A_u'^+B_u'^+}}
       {3\left(\overline{{A_u'^+}^2}+\overline{{A_w'^+}^2}\right)^3}
                                                                                                                                    \notag\\
&&
&{\color{blue}{+\left[\dfrac{4\overline{A_u'^+A_w'^+}^2\left(\overline{A_u'^+B_u'^+}+\overline{A_w'^+B_w'^+}\right)-2\overline{A_u'^+A_w'^+}\left(\overline{{A_u'^+}^2}+\overline{{A_w'^+}^2}\right)\left(\overline{A_u'^+B_w'^+}+\overline{A_w'^+B_u'^+}\right)}
              {3\left(\overline{{A_u'^+}^2}+\overline{{A_w'^+}^2}\right)^3}\right]}}\Vast)y^++O({y^+}^2)
                                                                                                                                    \notag
\end{alignat}
\begin{alignat}{8}
A\sim&27\dfrac{\overline{{A_u'^+}^2}\,\overline{{A_w'^+}^2}\,\overline{{B_v'^+}^2}-\overline{{A_w'^+}^2}\,\overline{A_u'^+B_v'^+}^2
               {\color{blue}{-\left[\overline{A_u'^+A_w'^+}^2\overline{{B_v'^+}^2}+\overline{{A_u'^+}^2}\,\overline{A_w'^+B_v'^+}^2-2\overline{A_u'^+A_w'^+}\,\overline{A_u'^+B_v'^+}\,\overline{A_w'^+B_v'^+}\right]}}}
              {\left(\overline{{A_u'^+}^2}+\overline{{A_w'^+}^2}\right)^3}{y^+}^2+O({y^+}^3)
                                                                                                                                    \notag
\end{alignat}
}

\caption{Asymptotic (as $y^+\to0$) expansions of $\tsr{b}$ \eqref{Eq_DTWT_s_DNSDepsij_ss_A_001a} and of its invariants \eqref{Eq_DTWT_s_DNSDepsij_ss_A_001b}, for general inhomogeneous incompressible \eqref{Eq_DTWT_s_epsijBs_ss_epsijTEq_001a}
         turbulent flow near a plane no-slip \eqrefsab{Eq_DTWT_s_AppendixABVSy+0_001}
                                                      {Eq_DTWT_s_AppendixABVSy+0_002}
         $xz$-wall (terms within square brackets {\color{blue}{$[\cdots]$}} are 3-D terms which are identically $=0$ for 2-D in-the-mean flow).}
\label{Tab_DTWT_s_DNSDepsij_ss_WA_001}
\rule{\textwidth}{.25pt}
\end{center}
\end{sidewaystable}
%
\begin{sidewaystable}
\vspace{4.7in}
\begin{center}
\rule{\textwidth}{.25pt}
\scalefont{.7}{
\begin{alignat}{8}
b_{\varepsilon_{xx}}&\sim& \dfrac{2\overline{{A_u'^+}^2}-\overline{{A_w'^+}^2}}
                                 {3\left(\overline{{A_u'^+}^2}+\overline{{A_w'^+}^2}\right)}
                    &   +&4\dfrac{\overline{{A_w'^+}^2}\,\overline{A_u'^+B_u'^+}-\overline{{A_u'^+}^2}\,\overline{A_w'^+B_w'^+}}
                                 {\left(\overline{{A_u'^+}^2}+\overline{{A_w'^+}^2}\right)^2}\,y^+
                    &   +&   O({y^+}^2)
                                                                                                                                    \notag\\
b_{\varepsilon_{xy}}&\sim&
                    &    &2\dfrac{\overline{A_u'^+B_v'^+}}
                                 {\overline{{A_u'^+}^2}+\overline{{A_w'^+}^2}}\,y^+
                    &   +&O({y^+}^2)
                                                                                                                                    \notag\\
b_{\varepsilon_{yy}}&\sim&-\tfrac{1}{3}
                    &    &
                    &   +&4\dfrac{\overline{{B_v'^+}^2}}
                                 {\overline{{A_u'^+}^2}+\overline{{A_w'^+}^2}}\,{y^+}^2
                    &   +&   O({y^+}^3)
                                                                                                                                    \notag\\
b_{\varepsilon_{yz}}&\sim&
                    &    &{\color{blue}{\vast[2\dfrac{\overline{A_w'^+B_v'^+}}
                                                     {\overline{{A_u'^+}^2}+\overline{{A_w'^+}^2}}\,y^+}}
                    &{\color{blue}{+}}
                         &{\color{blue}{O({y^+}^2)\vast]}}
                                                                                                                                    \notag\\
b_{\varepsilon_{zz}}&\sim& \dfrac{2\overline{{A_w'^+}^2}-\overline{{A_u'^+}^2}}
                                 {3\left(\overline{{A_u'^+}^2}+\overline{{A_w'^+}^2}\right)}
                    &    &-4\dfrac{\overline{{A_w'^+}^2}\,\overline{A_u'^+B_u'^+}-\overline{{A_u'^+}^2}\,\overline{A_w'^+B_w'^+}}
                                  {\left(\overline{{A_u'^+}^2}+\overline{{A_w'^+}^2}\right)^2}\,y^+
                    &   +& O({y^+}^2)
                                                                                                                                    \notag\\
b_{\varepsilon_{zx}}&\sim&{\color{blue}{\Bigg[\dfrac{\overline{A_u'^+A_w'^+}}
                                                    {\overline{{A_u'^+}^2}+\overline{{A_w'^+}^2}}}}
                    &   +&{\color{blue}{2\dfrac{\left(\overline{{A_u'^+}^2}+\overline{{A_w'^+}^2}\right)\left(\overline{A_u'^+B_w'^+}+\overline{A_w'^+B_u'^+}\right)
                                                -2\overline{A_u'^+A_w'^+}\left(\overline{A_u'^+B_u'^+}+\overline{A_w'^+B_w'^+}\right)}
                                               {\left(\overline{{A_u'^+}^2}+\overline{{A_w'^+}^2}\right)^2}\,y^+}}
                    &   +&{\color{blue}{ O({y^+}^2)\Bigg]}}
                                                                                                                                    \notag
\end{alignat}
\begin{alignat}{8}
\II{b_\varepsilon}\sim&-\dfrac{\overline{{A_u'^+}^2}^2-\overline{{A_u'^+}^2}\,\overline{{A_w'^+}^2}+\overline{{A_w'^+}^2}^2{\color{blue}{+\left[3\overline{A_u'^+A_w'^+}^2\right]}}}{3\left(\overline{{A_u'^+}^2}+\overline{{A_w'^+}^2}\right)^2}
&+&2\Vast(
2\dfrac{\left(\overline{{A_u'^+}^2}^2-\overline{{A_u'^+}^2}\,\overline{{A_w'^+}^2}\right)\overline{A_w'^+B_w'^+}+\left(\overline{{A_w'^+}^2}^2-\overline{{A_u'^+}^2}\,\overline{{A_w'^+}^2}\right)\overline{A_u'^+B_u'^+}}
       {\left(\overline{{A_u'^+}^2}+\overline{{A_w'^+}^2}\right)^3}
                                                                                                                                    \notag\\
&&
&{\color{blue}{+\left[\dfrac{4\overline{A_u'^+A_w'^+}^2\left(\overline{A_u'^+B_u'^+}+\overline{A_w'^+B_w'^+}\right)-2\overline{A_u'^+A_w'^+}\left(\overline{{A_u'^+}^2}+\overline{{A_w'^+}^2}\right)\left(\overline{A_u'^+B_w'^+}+\overline{A_w'^+B_u'^+}\right)}
              {\left(\overline{{A_u'^+}^2}+\overline{{A_w'^+}^2}\right)^3}\right]}}\Vast)y^++O({y^+}^2)
                                                                                                                                    \notag
\end{alignat}
\begin{alignat}{8}
\III{b_\varepsilon}\sim&\dfrac{2\left(\overline{{A_u'^+}^2}^2+\overline{{A_w'^+}^2}^2\right)-5\overline{{A_u'^+}^2}\,\overline{{A_w'^+}^2}{\color{blue}{+\left[9\overline{A_u'^+A_w'^+}^2\right]}}}{27\left(\overline{{A_u'^+}^2}+\overline{{A_w'^+}^2}\right)^2}
&-&2\Vast(
2\dfrac{\left(\overline{{A_u'^+}^2}^2-\overline{{A_u'^+}^2}\,\overline{{A_w'^+}^2}\right)\overline{A_w'^+B_w'^+}+\left(\overline{{A_w'^+}^2}^2-\overline{{A_u'^+}^2}\,\overline{{A_w'^+}^2}\right)\overline{A_u'^+B_u'^+}}
       {3\left(\overline{{A_u'^+}^2}+\overline{{A_w'^+}^2}\right)^3}
                                                                                                                                    \notag\\
&&
&{\color{blue}{+\left[\dfrac{4\overline{A_u'^+A_w'^+}^2\left(\overline{A_u'^+B_u'^+}+\overline{A_w'^+B_w'^+}\right)-2\overline{A_u'^+A_w'^+}\left(\overline{{A_u'^+}^2}+\overline{{A_w'^+}^2}\right)\left(\overline{A_u'^+B_w'^+}+\overline{A_w'^+B_u'^+}\right)}
              {3\left(\overline{{A_u'^+}^2}+\overline{{A_w'^+}^2}\right)^3}\right]}}\Vast)y^++O({y^+}^2)
                                                                                                                                    \notag
\end{alignat}
\begin{alignat}{8}
A_\varepsilon\sim&108\dfrac{\overline{{A_u'^+}^2}\,\overline{{A_w'^+}^2}\,\overline{{B_v'^+}^2}-\overline{{A_w'^+}^2}\,\overline{A_u'^+B_v'^+}^2
                            {\color{blue}{-\left[\overline{A_u'^+A_w'^+}^2\overline{{B_v'^+}^2}+\overline{{A_u'^+}^2}\,\overline{A_w'^+B_v'^+}^2-2\overline{A_u'^+A_w'^+}\,\overline{A_u'^+B_v'^+}\,\overline{A_w'^+B_v'^+}\right]}}}
                           {\left(\overline{{A_u'^+}^2}+\overline{{A_w'^+}^2}\right)^3}{y^+}^2+O({y^+}^3)
                                                                                                                                    \notag
\end{alignat}
}

\caption{Asymptotic (as $y^+\to0$) expansions of $\tsr{b_\varepsilon}$ \eqref{Eq_DTWT_s_DNSDepsij_ss_A_002a} and of its invariants \eqref{Eq_DTWT_s_DNSDepsij_ss_A_002b}, for general inhomogeneous incompressible \eqref{Eq_DTWT_s_epsijBs_ss_epsijTEq_001a}
         turbulent flow near a plane no-slip \eqrefsab{Eq_DTWT_s_AppendixABVSy+0_001}
                                                      {Eq_DTWT_s_AppendixABVSy+0_002}
         $xz$-wall (terms within square brackets {\color{blue}{$[\cdots]$}} are 3-D terms which are identically $=0$ for 2-D in-the-mean flow).}
\label{Tab_DTWT_s_DNSDepsij_ss_WA_002}
\rule{\textwidth}{.25pt}
\end{center}
\end{sidewaystable}
%
%
%
%
%
%
\subsection{Wall asymptotics}\label{DTWT_s_DNSDepsij_ss_WA}
%
%
%
%
%

Following the asymptotic (truncated Taylor-series) expansions \eqref{Eq_DTWT_s_AppendixABVSy+0_001}, the velocity components in the vicinity of the wall are expanded as \citep{Mansour_Kim_Moin_1988a}
\begin{subequations}
                                                                                                                                    \label{Eq_DTWT_s_DNSDepsij_ss_WA_001}
\begin{alignat}{6}
u'^+&\underset{y^+\to0}{\sim}&A_u'^+(x^+,z^+,t^+)\,y^+&+&B_u'^+(x^+,z^+,t^+)\,{y^+}^2+C_u'^+(x^+,z^+,t^+)\,{y^+}^3+\cdots
                                                                                                                                    \label{Eq_DTWT_s_DNSDepsij_ss_WA_001a}\\
v'^+&\underset{y^+\to0}{\sim}&                        &~&B_v'^+(x^+,z^+,t^+)\,{y^+}^2+C_v'^+(x^+,z^+,t^+)\,{y^+}^3+\cdots
                                                                                                                                    \label{Eq_DTWT_s_DNSDepsij_ss_WA_001b}\\
w'^+&\underset{y^+\to0}{\sim}&A_w'^+(x^+,z^+,t^+)\,y^+&+&B_w'^+(x^+,z^+,t^+)\,{y^+}^2+C_w'^+(x^+,z^+,t^+)\,{y^+}^3+\cdots
                                                                                                                                    \label{Eq_DTWT_s_DNSDepsij_ss_WA_001c}
\end{alignat}
\end{subequations}
to satisfy the no-slip conditon $u_i'(x,y^+=0,z,t)=0$ at the wall \eqref{Eq_DTWT_s_AppendixABVSy+0_002}, and the fluctuating continuity equation \eqref{Eq_DTWT_s_epsijBs_ss_epsijTEq_001a},
whose limit at the wall is $\lim_{y^+\to0}(\partial_{y^+} v'^+)=0\stackrel{\eqref{Eq_DTWT_s_AppendixABVSy+0_001}}{\implies}A_v'^+=0$, \ie\ $v'^+\propto {y^+}^2$ as $y^+\to0$ \eqref{Eq_DTWT_s_DNSDepsij_ss_WA_001b}.
Expansions \eqref{Eq_DTWT_s_DNSDepsij_ss_WA_001} of the near-wall fluctuating-velocity field are valid for general 3-D $xyz$-inhomogeneous turbulent incompressible flow near a plane no-slip wall.
Assuming \eqref{Eq_DTWT_s_DNSDepsij_ss_WA_001}, straightforward calculations yield the asymptotic expansions of $r_{ij}^+$ and $\varepsilon_{ij}^+$,
from which \parref{DTWT_s_AppendixABVSy+0_ss_ATsIs} are calculated the asymptotic expansions of $b_{ij}$ \tabref{Tab_DTWT_s_DNSDepsij_ss_WA_001},
$b_{\varepsilon_{ij}}$ \tabref{Tab_DTWT_s_DNSDepsij_ss_WA_002}, and of their invariants \tabrefsab{Tab_DTWT_s_DNSDepsij_ss_WA_001}
                                                                                                  {Tab_DTWT_s_DNSDepsij_ss_WA_002}.
Obviously \tabrefsab{Tab_DTWT_s_DNSDepsij_ss_WA_001}
                    {Tab_DTWT_s_DNSDepsij_ss_WA_002}
the wall-values of the anisotropy tensors $b_{ij}$ and $b_{\varepsilon_{ij}}$ are identical, and their wall-normal gradients are proportional with factor $2$, \viz
\begin{subequations}
                                                                                                                                    \label{Eq_DTWT_s_DNSDepsij_ss_WA_002}
\begin{alignat}{8}
b_{\varepsilon_{ij}}\underset{y^+\to0}{\sim}\left(b_{ij}\right)_w+2\left(\dfrac{\partial b_{ij}}{\partial y^+}\right)_w\,y^++O({y^+}^2)
                                                                                                                                    \label{Eq_DTWT_s_DNSDepsij_ss_WA_002a}
\end{alignat}
with fundamental differences occuring in the $O({y^+}^2)$-terms, those of $b_{\varepsilon_{ij}}$ containing the spatial gradients $\left(\nabla A_u'\right)^+$ and $\left(\nabla A_w'\right)^+$, contrary to those of $b_{ij}$.
Concerning the wall-normal diagonal components, $b_{yy}$ and $b_{\varepsilon_{yy}}$, whose linear $O(y^+)$ terms are $=0$, notice that \tabrefsab{Tab_DTWT_s_DNSDepsij_ss_WA_001}
                                                                                                                                                 {Tab_DTWT_s_DNSDepsij_ss_WA_002}
the wall-normal 2-derivatives are proportional with a factor 4
\begin{alignat}{8}
b_{\varepsilon_{yy}}\underset{y^+\to0}{\sim}\left(b_{yy}\right)_w+4\left(\tfrac{1}{2}\dfrac{\partial^2 b_{yy}}{\partial{y^+}^2}\right)_w\,{y^+}^2+O({y^+}^3)
                                                                                                                                    \label{Eq_DTWT_s_DNSDepsij_ss_WA_002b}
\end{alignat}
Relation \eqref{Eq_DTWT_s_DNSDepsij_ss_WA_002a} carries over \tabrefsab{Tab_DTWT_s_DNSDepsij_ss_WA_001}
                                                                       {Tab_DTWT_s_DNSDepsij_ss_WA_002}
to the anisotropy invariants \eqrefsab{Eq_DTWT_s_DNSDepsij_ss_A_001b}
                                      {Eq_DTWT_s_DNSDepsij_ss_A_002b}
\begin{alignat}{8}
 \II{b_\varepsilon}&\underset{y^+\to0}{\sim}&\left( \II{b}\right)_w&&+&2\left(\dfrac{\partial \II{b}}{\partial y^+}\right)_w\,y^+&+O({y^+}^2)
                                                                                                                                    \label{Eq_DTWT_s_DNSDepsij_ss_WA_002c}\\
\III{b_\varepsilon}&\underset{y^+\to0}{\sim}&\left(\III{b}\right)_w&&+&2\left(\dfrac{\partial\III{b}}{\partial y^+}\right)_w\,y^+&+O({y^+}^2)
                                                                                                                                    \label{Eq_DTWT_s_DNSDepsij_ss_WA_002d}
\end{alignat}
\end{subequations}
Consistent with the 2-C componentality of both $r_{ij}$ \eqref{Eq_DTWT_s_DNSDepsij_ss_A_001a} and $\varepsilon_{ij}$ \eqref{Eq_DTWT_s_DNSDepsij_ss_A_002a}
at the wall \citep{Launder_Shima_1989a,
                   Hanjalic_1994a,
                   Craft_Launder_2001a},
the wall-values of the corresponding Lumley's flatness parameters \eqrefsab{Eq_DTWT_s_DNSDepsij_ss_A_001b}
                                                                           {Eq_DTWT_s_DNSDepsij_ss_A_002b}
$A_w=\left(A_\varepsilon\right)_w=0$ \citep{Lumley_1978a,
                                            Simonsen_Krogstad_2005a}.
Furthermore, both $A$ and $A_\varepsilon$ tend to $0$ quadratically as $y^+\to0$, and their wall-normal 2-derivatives are proportional by a factor 4 \tabrefsab{Tab_DTWT_s_DNSDepsij_ss_WA_001}
                                                                                                                                                           {Tab_DTWT_s_DNSDepsij_ss_WA_002}
\begin{subequations}
                                                                                                                                    \label{Eq_DTWT_s_DNSDepsij_ss_WA_003}
\begin{alignat}{8}
 A            &\underset{y^+\to0}{\sim}&O({y^+}^2&)
                                                                                                                                    \label{Eq_DTWT_s_DNSDepsij_ss_WA_003a}\\
 A_\varepsilon&\underset{y^+\to0}{\sim}&4\left(\tfrac{1}{2}\dfrac{\partial^2A}{\partial{y^+}^2}\right)_w\,{y^+}^2&+O({y^+}^3)
                                                                                                                                    \label{Eq_DTWT_s_DNSDepsij_ss_WA_003b}
\end{alignat}
because, as can be verified by straightforward calculation, the constant and linear terms in the expansion of the invariants $\{\II{b},\III{b},\II{b_\varepsilon},\III{b_\varepsilon}\}$ \tabrefsab{Tab_DTWT_s_DNSDepsij_ss_WA_001}
                                                                                                                                                                                                   {Tab_DTWT_s_DNSDepsij_ss_WA_002},
cancel out
\begin{alignat}{8}
9&&\left( \II{b}\right)_w                              &+27&\left(\III{b}\right)_w                              &=9&\left( \II{b_\varepsilon}\right)_w                              &+27&\left(\III{b_\varepsilon}\right)_w                              &=&-1
                                                                                                                                    \label{Eq_DTWT_s_DNSDepsij_ss_WA_003c}\\
9&&\left(\dfrac{\partial \II{b}}{\partial y^+}\right)_w&+27&\left(\dfrac{\partial\III{b}}{\partial y^+}\right)_w&=9&\left(\dfrac{\partial \II{b_\varepsilon}}{\partial y^+}\right)_w&+27&\left(\dfrac{\partial\III{b_\varepsilon}}{\partial y^+}\right)_w&=&0
                                                                                                                                    \label{Eq_DTWT_s_DNSDepsij_ss_WA_003d}
\end{alignat}
\end{subequations}
Of course, calculations retaining the $O(y^{+2})$-terms in the expansions of the invariants were required to obtain the result \eqref{Eq_DTWT_s_DNSDepsij_ss_WA_003d}.

Regarding $b_{ij}$ \tabref{Tab_DTWT_s_DNSDepsij_ss_WA_001}, the new results in the paper, with reference to \citet[Tab. 3, p. 191]{Hanjalic_1994a} are the asymptotic expansions of the invariants $\II{b}$ and $\III{b}$,
and in particular of the flatness parametr $A$. Corresponding expansions were obtained \tabref{Tab_DTWT_s_DNSDepsij_ss_WA_002} for $b_{\varepsilon_{ij}}$ and for its invarianrs ($\II{b_{\varepsilon}}$, $\III{b_{\varepsilon}}$, $A_\varepsilon$).
With regard to the relations between $b_{ij}$ and $b_{\varepsilon_{ij}}$, \citet[p. 32]{Mansour_Kim_Moin_1988a} have "{\em point{\em(ed)} out that close to the wall, Taylor-series expansions
of $b_{\varepsilon_{ij}}$ and $b_{ij}$ show that they are
equal only up to $O(y^+)$}"; this remark clearly implies $(b_{\varepsilon_{ij}})_w=(b_{ij})_w$. The present results \eqrefsab{Eq_DTWT_s_DNSDepsij_ss_WA_002}{Eq_DTWT_s_DNSDepsij_ss_WA_003} show the proportionality relation 
$(\partial_{y^+} b_{ij})_w\stackrel{\eqref{Eq_DTWT_s_DNSDepsij_ss_WA_002a}}{=}\tfrac{1}{2}(\partial_{y^+} b_{\varepsilon_{ij}})_w$  of the wall-normal gradients and the corresponding relation between the wall-normal gradients of the invariants
$(\partial_{y^+} \II{b_{\varepsilon}})_w\stackrel{\eqref{Eq_DTWT_s_DNSDepsij_ss_WA_002c}}{=}2(\partial_{y^+} \II{b})_w$ and 
$(\partial_{y^+} \III{b_{\varepsilon}})_w\stackrel{\eqref{Eq_DTWT_s_DNSDepsij_ss_WA_002d}}{=}2(\partial_{y^+} \III{b})_w$ on the one hand, and the quadratic asymptotic behaviour of the flatness parameters
$A_\varepsilon\stackrel{\eqref{Eq_DTWT_s_DNSDepsij_ss_WA_003}}{\underset{y^+\to0}{\sim}}4 A + O(y^{+3})\stackrel{\eqref{Eq_DTWT_s_DNSDepsij_ss_WA_003a}}{\underset{y^+\to0}{\propto}}O(y^{+2})$ on the other hand.

%
%
%
%
%
\subsection{$Re_{\tau_w}$-influence on anisotropy}\label{DTWT_s_DNSDepsij_ss_RetauwIA}
%
%
%
%
%

The evolution of mean-flow and turbulence structure with $Re_{\tau_w}$ \eqref{Eq_DTWT_s_AppendixABVSy+0_ss_WUs_001g} is central in wall-turbulence  research \citep{Marusic_McKeon_Monkewitz_Nagib_Smits_Sreenivasan_2010a,
                                                                                                                                                                    Marusic_Monty_Hultmark_Smits_2013a,
                                                                                                                                                                    Kim_2012a}
and \tsn{DNS} of turbulent plane channel flow at progressively higher $Re_{\tau_w}$ \citep{Moser_Kim_Mansour_1999a,
                                                                                           Hoyas_Jimenez_2006a,
                                                                                           Hoyas_Jimenez_2008a,
                                                                                           Bernardini_Pirozzoli_Orlandi_2014a,
                                                                                           Lee_Moser_2015a}
contribute towards answering several open questions on very-high-$Re$ asymptotics.
Turbulence structure is generally represented by nondimensional ratios of turbulent quantities \citep{Bradshaw_1967a}.
In this respect, the anisotropy tensors $b_{ij}$ \eqref{Eq_DTWT_s_DNSDepsij_ss_A_001a} and $b_{\varepsilon_{ij}}$ \eqref{Eq_DTWT_s_DNSDepsij_ss_A_002a}, and their \tsn{AIM}s \citep{Lumley_Newman_1977a,
                                                                                                                                                                                     Lumley_1978a,
                                                                                                                                                                                     Lee_Reynolds_1987a,
                                                                                                                                                                                     Simonsen_Krogstad_2005a}
are particularly useful in understanding the differences in behaviour between the Reynolds-stresses $r_{ij}$ \eqref{Eq_DTWT_s_DNSDepsij_ss_A_001a} and their dissipation-rates $\varepsilon_{ij}$ \eqref{Eq_DTWT_s_DNSDepsij_ss_A_002a},
across the channel (in the wall-normal direction $y$) and with varying Reynolds number $Re_{\tau_w}$ \eqref{Eq_DTWT_s_AppendixABVSy+0_ss_WUs_001g}.
\begin{figure}
\begin{center}
\begin{picture}(450,450)
\put(0,0){\includegraphics[angle=0,width=380pt,bb=73 178 526 705]{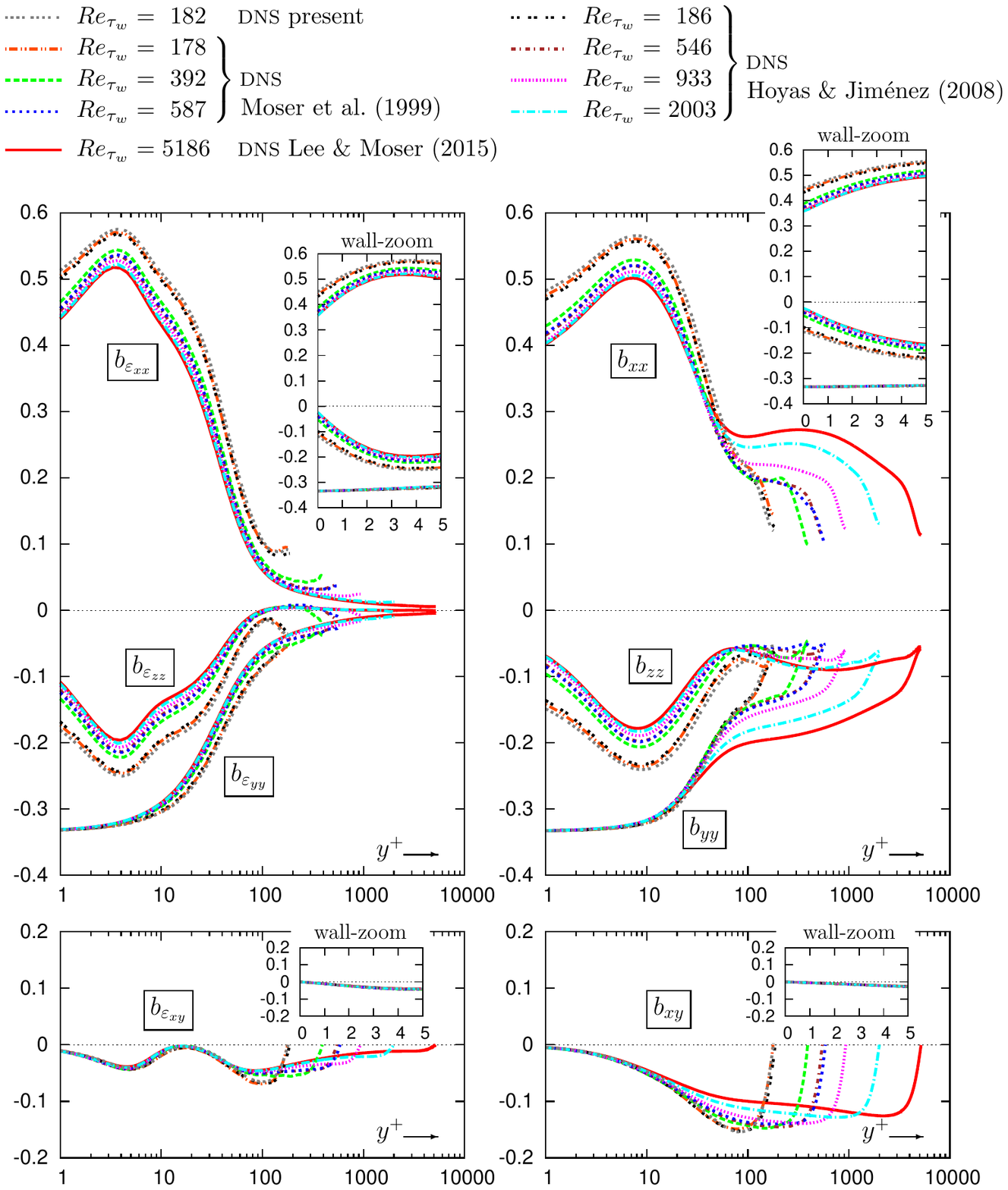}}
\end{picture}
\end{center}
\caption{Anisotropy \eqrefsab{Eq_DTWT_s_DNSDepsij_ss_A_001}
                             {Eq_DTWT_s_DNSDepsij_ss_A_002}
of the normal ($b_{xx}$, $b_{yy}$, $b_{zz}$) and shear ($b_{xy}$) Reynolds-stresses,
and of the corresponding dissipation-rates ($b_{\varepsilon_{xx}}$, $b_{\varepsilon_{yy}}$, $b_{\varepsilon_{zz}}$, $b_{\varepsilon_{xy}}$),
from existing \citep{Moser_Kim_Mansour_1999a,
                     delAlamo_Jimenez_2003a,
                     Hoyas_Jimenez_2006a,
                     Hoyas_Jimenez_2008a,
                     Lee_Moser_2015a}
\tsn{DNS} computations of turbulent plane channel flow, in the range $Re_{\tau_w}\in[180,5200]$,
plotted against the inner-scaled \eqref{Eq_DTWT_s_AppendixABVSy+0_ss_WUs_001c} wall-distance $y^+$ (logscale and linear wall-zoom).}
\label{Fig_DTWT_s_DNSDepsij_ss_RetauwIA_sss_ATs_001}
\end{figure}
%
\begin{figure}
\begin{center}
\begin{picture}(450,390)
\put(0,0){\includegraphics[angle=0,width=380pt,bb=58 280 526 744]{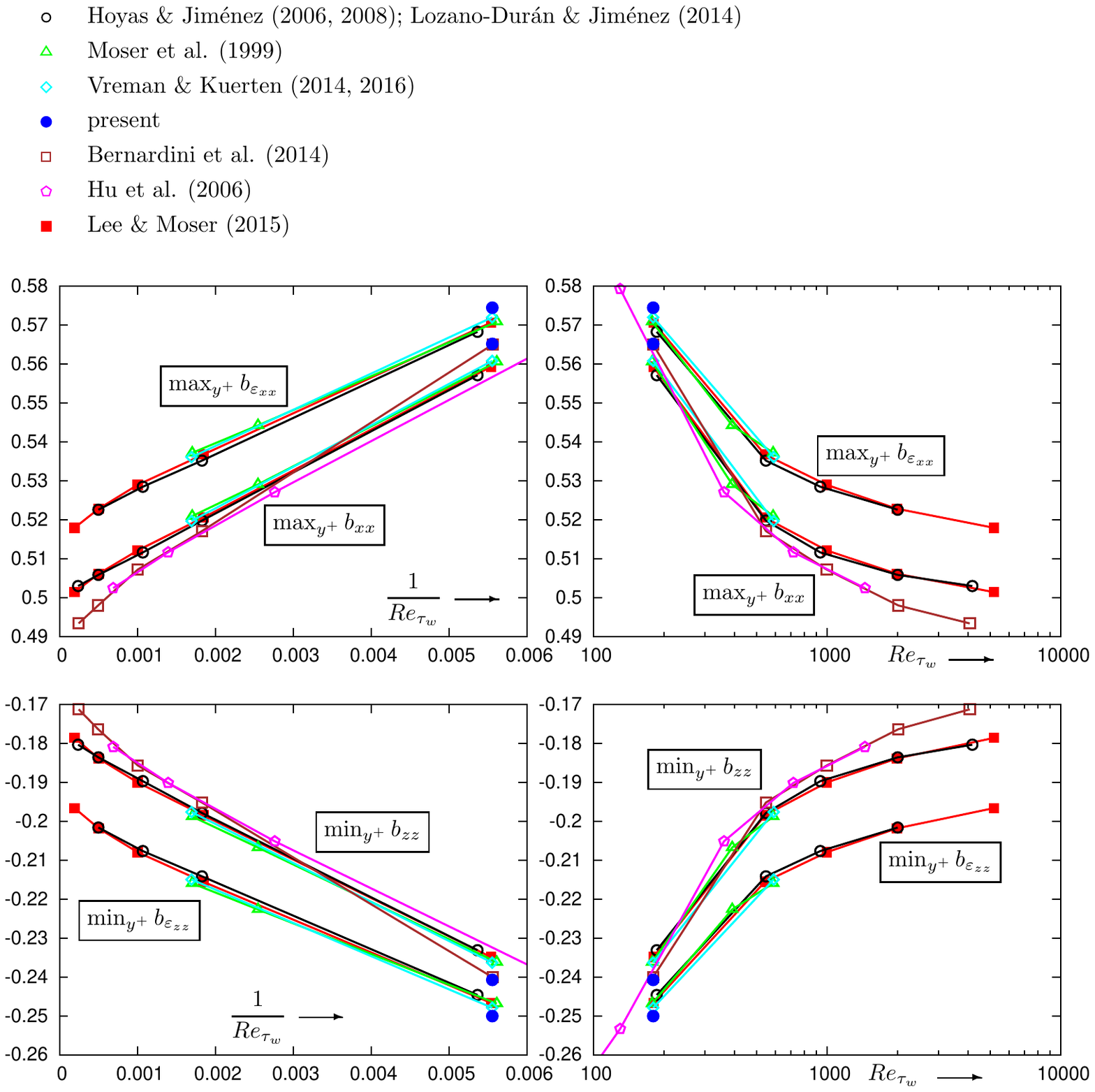}}
\end{picture}
\end{center}
\caption{Near-wall peaks of the steamwise ($\max_{y^+}b_{xx}$, $\max_{y^+}b_{\varepsilon_{xx}}$) and spanwise ($\min_{y^+}b_{zz}$, $\min_{y^+}b_{\varepsilon_{zz}}$)
components of the anisotropy-tensors of the Reynolds-stresses $\tsr{b}$ \eqref{Eq_DTWT_s_DNSDepsij_ss_A_001} and of the corresponding dissipation-rates $\tsr{b_\varepsilon}$ \eqref{Eq_DTWT_s_DNSDepsij_ss_A_002},
as a function of the friction Reynolds number \eqref{Eq_DTWT_s_AppendixABVSy+0_ss_WUs_001g} $Re_{\tau_w}$ (logscale) and of its inverse $Re_{\tau_w}^{-1}$ (linear scale),
from existing \citep{Moser_Kim_Mansour_1999a,
                     Hu_Morfey_Sandham_2006a,
                     Hoyas_Jimenez_2006a,
                     Hoyas_Jimenez_2008a,
                     LozanoDuran_Jimenez_2014a,
                     Bernardini_Pirozzoli_Orlandi_2014a,
                     Vreman_Kuerten_2014b,
                     Vreman_Kuerten_2016a,
                     Lee_Moser_2015a}
\tsn{DNS} computations of turbulent plane channel flow, in the range $Re_{\tau_w}\in[100,5200]$.}
\label{Fig_DTWT_s_DNSDepsij_ss_RetauwIA_sss_ATs_002}
\end{figure}

%
\subsubsection{Anisotropy tensors}\label{DTWT_s_DNSDepsij_ss_RetauwIA_sss_ATs}
%

The strong anisotropy of the Reynolds-stresses in wall-turbulence is related to the anisotropy of the dissipation-rate tensor $\varepsilon_{ij}$ \citep{Gerolymos_Lo_Vallet_Younis_2012a},
and it is established that the behaviour of the anisotropy tensor of the Reynolds-stresses $b_{ij}$ differs from that of the anisotropy-tensor of the dissipation-rate $b_{\varepsilon_{ij}}$ \citep{Launder_Reynolds_1983a,
                                                                                                                                                                                                     Lai_So_1990a,
                                                                                                                                                                                                     Jovanovic_Ye_Durst_1995a,
                                                                                                                                                                                                     Jakirlic_Hanjalic_2002a}.
This is obvious by examining \figref{Fig_DTWT_s_DNSDepsij_ss_RetauwIA_sss_ATs_001}
existing \citep{Moser_Kim_Mansour_1999a,
                delAlamo_Jimenez_2003a,
                Hoyas_Jimenez_2006a,
                Hoyas_Jimenez_2008a,
                Lee_Moser_2015a}
\tsn{DNS} results which include budgets of $r_{ij}$-transport (and hence data for the components of $\varepsilon_{ij}$), covering the range $Re_{\tau_w}\in[178,5186]$.
Recall that \parref{DTWT_s_DNSDepsij_ss_WA}, at the wall ($y^+=0$), the anisotropy tensors $b_{ij}$ and $b_{\varepsilon_{ij}}$ are identical \citep[p. 32]{Mansour_Kim_Moin_1988a},
but the wall-normal gradient of $b_{\varepsilon_{ij}}$ is exactly twice that of $b_{ij}$ \eqref{Eq_DTWT_s_DNSDepsij_ss_WA_002a}.
This explains the near-wall ($y^+<10$) evolution of the streamwise ($b_{xx}$, $b_{\varepsilon_{xx}}$) and spanwise ($b_{zz}$, $b_{\varepsilon_{zz}}$) components of the anisotropy tensors \figref{Fig_DTWT_s_DNSDepsij_ss_RetauwIA_sss_ATs_001},
\viz\ that $\{b_{\varepsilon_{xx}},\abs{b_{\varepsilon_{zz}}}\}$ increase much faster with $y^+$, and reach their maxima at $y^+\in[3,4]$, nearer the wall compared to
$\{b_{xx},\abs{b_{zz}}\}$ which reach their maxima at $y^+\in[7,9]$. These near-wall maxima of $\{b_{xx},\abs{b_{zz}},b_{\varepsilon_{xx}},\abs{b_{\varepsilon_{zz}}}\}$ are also global maxima across the channel \figref{Fig_DTWT_s_DNSDepsij_ss_RetauwIA_sss_ATs_001}.
Notice that at fixed $Re_{\tau_w}$ the maximum of the spanwise component occurs at slightly higher $y^+$ than the maximum of the streamwise component, both for $b_{ij}$ 
and $b_{\varepsilon_{ij}}$ \figref{Fig_DTWT_s_DNSDepsij_ss_RetauwIA_sss_ATs_001}. Furthermore, at fixed $Re_{\tau_w}$, the values of the maxima are quite close ($\max_y b_{xx}\approxeq\max_y b_{\varepsilon_{xx}}$ and
$\max_y \abs{b_{zz}}\approxeq\max_y \abs{b_{\varepsilon_{zz}}}$) but $\varepsilon_{ij}$ is slightly more anisotropic than $r_{ij}$  \figref{Fig_DTWT_s_DNSDepsij_ss_RetauwIA_sss_ATs_001}.
In the near-wall region ($y^+\lessapprox 10$), the anisotropy of the streamwise ($b_{xx}$, $b_{\varepsilon_{xx}}$) and spanwise ($b_{zz}$, $b_{\varepsilon_{zz}}$) components decreases with increasing $Re_{\tau_w}$, first quite sharply
($Re_{\tau_w}<400$) and then more slowly \figref{Fig_DTWT_s_DNSDepsij_ss_RetauwIA_sss_ATs_001}. Careful examination of the values of the maxima \vs\ $Re_{\tau_w}$ \figref{Fig_DTWT_s_DNSDepsij_ss_RetauwIA_sss_ATs_002}
suggests that they probably tend to $Re_{\tau_w}$-independent values as $Re_{\tau_w}$ increases, although these are not reached yet at the highest available $Re_{\tau_w}\approxeq 5186$ \tsn{DNS} of \citet{Lee_Moser_2015a}.
To alleviate the bias of the computational grid, the values of the extrema were determined by fitting a degree-4 interpolating polynomial $p_I(y^+)$ around the on-grid extremum (2 neighbours on each side)
and solving $p_I'(y^+)=0$ by Newton iteration. This generally had no influence on the value of the maximum, but did smooth out variations of its location.
Plots \figref{Fig_DTWT_s_DNSDepsij_ss_RetauwIA_sss_ATs_002} of the near-wall peaks of the steamwise ($\max_{y^+}b_{xx}$, $\max_{y^+}b_{\varepsilon_{xx}}$) and spanwise ($\min_{y^+}b_{zz}$, $\min_{y^+}b_{\varepsilon_{zz}}$)
components \vs\ $Re_{\tau_w}$ (logscale) hint that asymptotic limits are approached, and this is further verified by plotting the data against $Re_{\tau_w}^{-1}$ \figref{Fig_DTWT_s_DNSDepsij_ss_RetauwIA_sss_ATs_002}.
More sets of data are available for $r_{ij}$ than for $\varepsilon_{ij}$, and 2 slightly distinct curves appear for the $b_{ij}$-extrema \figref{Fig_DTWT_s_DNSDepsij_ss_RetauwIA_sss_ATs_002},
but this only affects the precise value of the asymptotic limits as $Re_{\tau_w}^{-1}\to0$ ($\sim2\%$ scatter for $\max_{y^+}b_{xx}$ and $\sim5\%$ scatter for $\min_{y^+}b_{zz}$).
The anisotropy of the wall-normal component is of course highest at the 2-C wall $(b_{yy})_w=(b_{\varepsilon_{yy}})_w=-\tfrac{1}{3}$ and then  becomes less anisotropic with increasing $y^+$ \figref{Fig_DTWT_s_DNSDepsij_ss_RetauwIA_sss_ATs_001}.
Obviously, the coefficient $\varepsilon_w^{+-1}\overline{B_v'^{+2}}$ \tabrefsab{Tab_DTWT_s_DNSDepsij_ss_WA_001}{Tab_DTWT_s_DNSDepsij_ss_WA_002} of the leading quadratic term in \eqref{Eq_DTWT_s_DNSDepsij_ss_WA_002b} slightly increases with increasing $Re_{\tau_w}$
\figref{Fig_DTWT_s_DNSDepsij_ss_RetauwIA_sss_ATs_001}. Finally, regarding the shear components, both $b_{xy}$ and $b_{\varepsilon_{xy}}$ seem $Re_{\tau_w}$-independent near the wall ($y^+\lessapprox 10$).

In contrast to the near-wall region ($y^+\lessapprox 10$), which is dominated by the wall-asymptotic relations \parref{DTWT_s_DNSDepsij_ss_WA}, the behaviour of $b_{\varepsilon_{ij}}$ is completely different from that of $b_{ij}$ at higher $y^+$
\figref{Fig_DTWT_s_DNSDepsij_ss_RetauwIA_sss_ATs_001}. The anisotropy of the streamwise Reynolds-stress component $b_{xx}$ forms, with increasing $Re_{\tau_w}$, a plateau corresponding to the log-region of the mean-velocity profile
\citep{Coles_1956a} followed by a sharp decrease in the wake-region.
The level of anisotropy $b_{xx}$ in the log-region increases with $Re_{\tau_w}$ but may be reaching an asymptotic limit, although simulations at higher $Re_{\tau_w}$ are needed to ascertain this point. 
The spanwise component $b_{zz}$ becomes more isotropic with increasing $y^+$ in the buffer layer ($y^+\in [10,100]$), followed by  a slight increase in anisotropy in the log-region \figref{Fig_DTWT_s_DNSDepsij_ss_RetauwIA_sss_ATs_001}, while the wall-normal component $b_{yy}$
becomes more isotropic with increasing $y^+$, quite sharply in the buffer-layer and more progressively in the log-region. 
Interestingly, both $b_{yy}$ and $b_{zz}$ also seem to be forming a log-region plateau with increasing $Re_{\tau_w}$, but simulations at higher $Re_{\tau_w}$
are required to verify this trend. Both components $b_{yy}$ and $b_{zz}$ become more anisotropic in the log-region with increasing $Re_{\tau_w}$ \figref{Fig_DTWT_s_DNSDepsij_ss_RetauwIA_sss_ATs_001}.
Finally, in the wake-region, both $b_{yy}$ and $b_{zz}$ tend quite sharply to a common centerline value \figref{Fig_DTWT_s_DNSDepsij_ss_RetauwIA_sss_ATs_001}.

Contrary to $b_{ij}$, the diagonal components of $b_{\varepsilon_{ij}}$ tend quasi-monotonically to a near-isotropic (but not exactly isotropic) state at the centerline, first steeply in the buffer-layer and then more gradually in the 
log-region \figref{Fig_DTWT_s_DNSDepsij_ss_RetauwIA_sss_ATs_001}. The low-$Re_{\tau_w}$ data exhibit a clear albeit slight increase in the anisotropy of $\{b_{\varepsilon_{xx}},b_{\varepsilon_{yy}},b_{\varepsilon_{zz}}\}$ near the centerline which becomes less pronounced
with increasing $Re_{\tau_w}$. Unlike $b_{zz}$, $b_{\varepsilon_{zz}}$ reaches a near-0 value in the log-region \figref{Fig_DTWT_s_DNSDepsij_ss_RetauwIA_sss_ATs_001}. 

The shear components also behave very differently. The Reynolds-stress anisotropy $b_{xy}$ 
becomes  more anisotropic from the wall up to the beginning of the wake-region, before sharply going to 0 (exactly, because of the symmetry condition $\overline{u'v'}\big|_\tsn{CL}=0$) at the centerline.
 The shear component $b_{\varepsilon_{xy}}$ is practically $Re_{\tau_w}$-independent up to the end of the buffer-layer ($y^+\lessapprox 100$),
with a wavy shape reaching 0 near $y^+\approxeq 20$ \figref{Fig_DTWT_s_DNSDepsij_ss_RetauwIA_sss_ATs_001}, then being relatively flat in the log-region.  
\begin{figure}
\begin{center}
\begin{picture}(450,545)
\put(0,-5){\includegraphics[angle=0,width=365pt,bb=73 84 526 759]{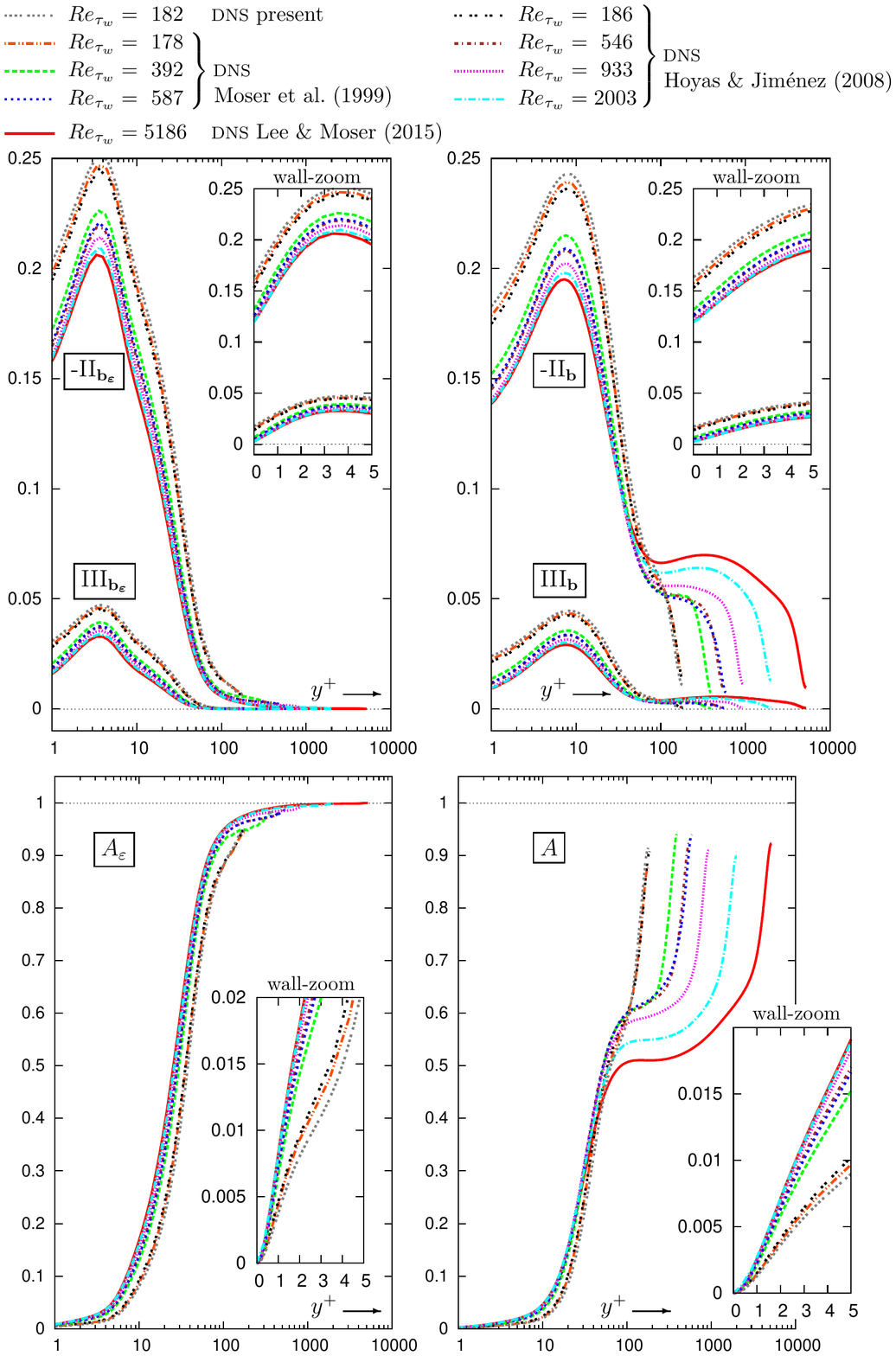}}
\end{picture}
\end{center}
\caption{Invariants of the anisotropy-tensors \eqrefsab{Eq_DTWT_s_DNSDepsij_ss_A_001}
                                                       {Eq_DTWT_s_DNSDepsij_ss_A_002}
of the Reynolds-stresses $\tsr{b}$ ($\II{b}$, $\III{b}$)
and of the corresponding dissipation-rates $\tsr{b_\varepsilon}$ ($\II{b_\varepsilon}$, $\III{b_\varepsilon}$),
and \citeauthor{Lumley_1978a}'s (\citeyear{Lumley_1978a}) flatness parameters
$A:=1+27\III{b}+\II{b}$ and $A_\varepsilon:=1+27\III{b_\varepsilon}+\II{b_\varepsilon}$),
from existing \citep{Moser_Kim_Mansour_1999a,
                     delAlamo_Jimenez_2003a,
                     Hoyas_Jimenez_2006a,
                     Hoyas_Jimenez_2008a,
                     Lee_Moser_2015a}
\tsn{DNS} computations of turbulent plane channel flow, in the range $Re_{\tau_w}\in[180,5200]$,
plotted against the inner-scaled \eqref{Eq_DTWT_s_AppendixABVSy+0_ss_WUs_001c} wall-distance $y^+$ (logscale and linear wall-zoom).}
\label{Fig_DTWT_s_DNSDepsij_ss_RetauwIA_sss_AIs_001}
\end{figure}

%
\subsubsection{Anisotropy invariants}\label{DTWT_s_DNSDepsij_ss_RetauwIA_sss_AIs}
%

More precise information on the influence of $Re_{\tau_w}$ \eqref{Eq_DTWT_s_AppendixABVSy+0_ss_WUs_001g} on the anisotropy of $r_{ij}$ and $\varepsilon_{ij}$ can be obtained by considering 
the invariants $\{\III{b},-\II{b}\}$ \eqref{Eq_DTWT_s_DNSDepsij_ss_A_001b} and $\{\III{b_\varepsilon},-\II{b_\varepsilon}\}$ \eqref{Eq_DTWT_s_DNSDepsij_ss_A_002b}
whose locus lies inside \citeauthor{Lumley_1978a}'s (\citeyear{Lumley_1978a}) realizability triangle \citep[Figs.~17 and 18, p.~32]{Mansour_Kim_Moin_1988a}. As the intervals
of possible values \citep{Lumley_1978a,
                          Simonsen_Krogstad_2005a}
\begin{subequations}
                                                                                                                                    \label{Eq_DTWT_s_DNSDepsij_ss_RetauwIA_001}
\begin{alignat}{8}
\II{b} &\in&[&-\tfrac{1}{3}   &,& 0                   &]           &\ni& \II{b_\varepsilon}    
                                                                                                                                    \label{Eq_DTWT_s_DNSDepsij_ss_RetauwIA_001a}\\
\III{b}&\in&[&-\tfrac{1}{108} &,& \tfrac{2}{27}      & ]           &\ni& \III{b_\varepsilon} 
                                                                                                                                    \label{Eq_DTWT_s_DNSDepsij_ss_RetauwIA_001b}\\
A      &\in&[& \quad        0 &,& 1                  & ]           &\ni& A_\varepsilon
                                                                                                                                    \label{Eq_DTWT_s_DNSDepsij_ss_RetauwIA_001c}
\end{alignat}
\end{subequations}
are rather limited, the invariants are quite sensitive indicators of anisotropy, and this sensitivity is visible in the influence of $Re_{\tau_w}$ on the near-wall peaks of the invariants \figref{Fig_DTWT_s_DNSDepsij_ss_RetauwIA_sss_AIs_001}.
The evolution of the invariants with $y^+$ and $Re_{\tau_w}$ \figref{Fig_DTWT_s_DNSDepsij_ss_RetauwIA_sss_AIs_001} is very similar to what was observed for the diagonal components of the anisotropy tensors \figref{Fig_DTWT_s_DNSDepsij_ss_RetauwIA_sss_ATs_001}.
Near the wall, $\{\III{b_\varepsilon},-\II{b_\varepsilon}\}$ increase faster with $y^+$ compared to $\{\III{b},-\II{b}\}$, in line with \eqrefsab{Eq_DTWT_s_DNSDepsij_ss_WA_002c}
                                                                                                                                                 {Eq_DTWT_s_DNSDepsij_ss_WA_002d},
reaching their maxima at $y^+\in[3,4]$ compared to $y^+\in[7,9]$ \figref{Fig_DTWT_s_DNSDepsij_ss_RetauwIA_sss_AIs_001}, $\varepsilon_{ij}$ being slightly more anisotropic than $r_{ij}$.
Near the wall, $A_\varepsilon$ increases much faster with $y^+$ compared to $A$ \figref{Fig_DTWT_s_DNSDepsij_ss_RetauwIA_sss_AIs_001}, in line with \eqrefsab{Eq_DTWT_s_DNSDepsij_ss_WA_003a}
                                                                                                                                                     {Eq_DTWT_s_DNSDepsij_ss_WA_003b}.
There is noticeable $Re_{\tau_w}$-influence for $Re_{\tau_w}<400$ both in wall values and in rate-of-increase with $y^+$ for all of the invariants \figref{Fig_DTWT_s_DNSDepsij_ss_RetauwIA_sss_AIs_001}.
On the other hand, with increasing $Re_{\tau_w}$, it would seem that an asymptotic state is approached
in the near-wall region, including the buffer-layer ($y^+\lessapprox 100$), although \tsn{DNS} at higher $Re_{\tau_w}$ are still required to fully substantiate this observation. Regarding the invariants of $r_{ij}$,
$\{-\II{b},\III{b},A\}$, a plateau appears with increasing $Re_{\tau_w}$ \figref{Fig_DTWT_s_DNSDepsij_ss_RetauwIA_sss_AIs_001}, marking the log-region of the mean-velocity profile \citep{Lee_Moser_2015a}, but again \tsn{DNS} at higher $Re_{\tau_w}$ are needed
to determine whether a $Re_{\tau_w}$-asymptotic value of the level of this plateau exists.
On the contrary, the invariants of $\varepsilon_{ij}$ $\{-\II{b_\varepsilon},\III{b_\varepsilon},A_\varepsilon\}$ vary monotonically from the wall to centerline, and seem to approach $Re_{\tau_w}$-independent $y^+$-distributions with increasing $Re_{\tau_w}$.

%
\subsubsection{AIM at the wall and at the centerline}\label{DTWT_s_DNSDepsij_ss_RetauwIA_sss_AIMWC}
%

\begin{figure}
\begin{center}
\begin{picture}(450,390)
\put(0,0){\includegraphics[angle=0,width=380pt,bb=58 280 526 748]{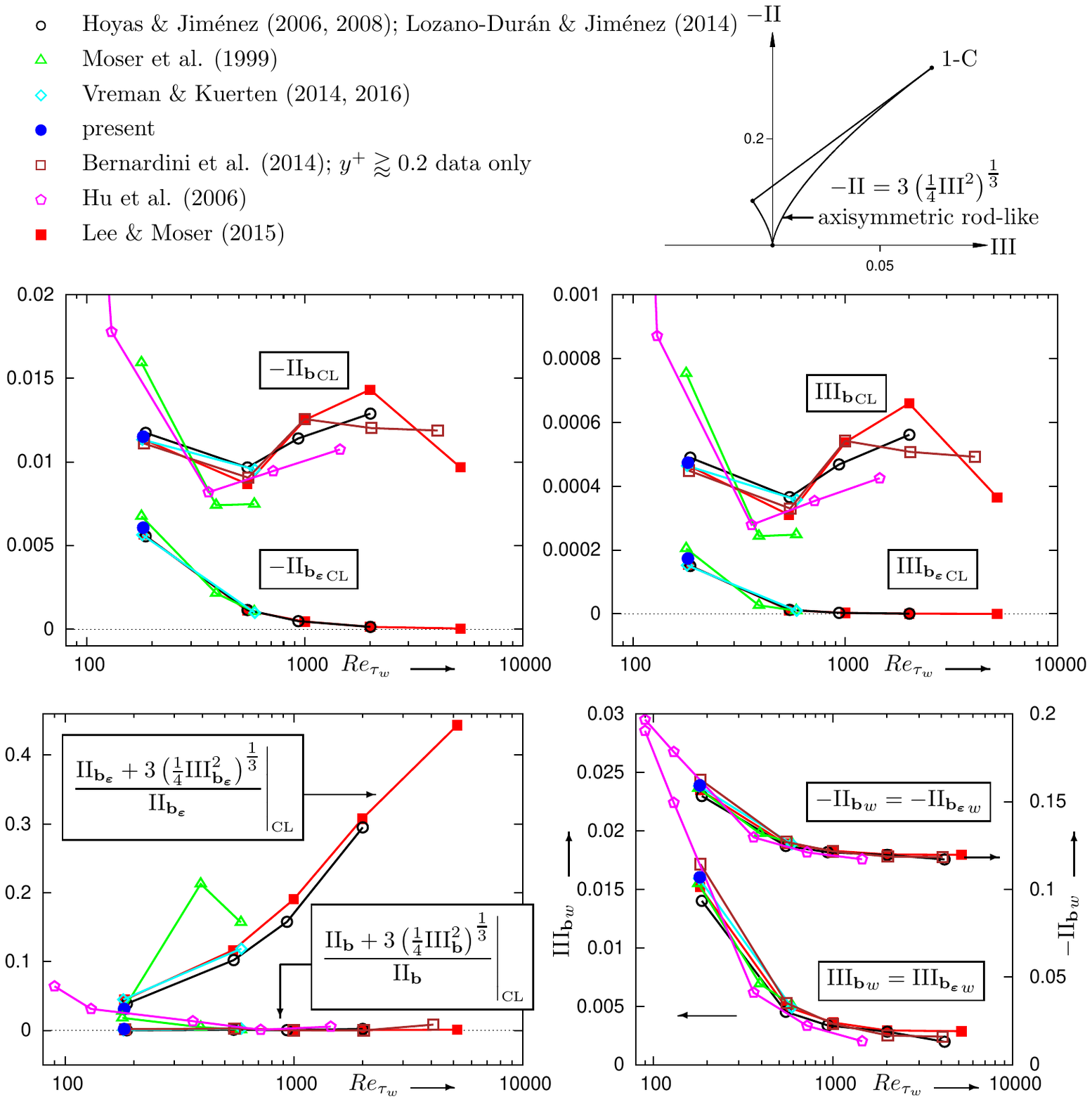}}
\end{picture}
\end{center}
\caption{Invariants of the anisotropy-tensors \eqrefsab{Eq_DTWT_s_DNSDepsij_ss_A_001}
                                                       {Eq_DTWT_s_DNSDepsij_ss_A_002}
of the Reynolds-stresses $\tsr{b}$ ($\II{b}$, $\III{b}$)
and of the corresponding dissipation-rates $\tsr{b_\varepsilon}$ ($\II{b_\varepsilon}$, $\III{b_\varepsilon}$), at the wall $(\cdot)_w$ and at channel centerline $(\cdot)_\tsn{CL}$,
and axisymmetry-diagnostic $\II{b}^{-1}\Big(\II{b}+3(\tfrac{1}{4}\III{b})^{\tfrac{1}{3}}\Big)$ and $\II{b_\varepsilon}^{-1}\Big(\II{b_\varepsilon}+3(\tfrac{1}{4}\III{b_\varepsilon})^{\tfrac{1}{3}}\Big)$,
at the centerline $(\cdot)_\tsn{CL}$, as a function of the friction Reynolds number $Re_{\tau_w}$ \eqref{Eq_DTWT_s_AppendixABVSy+0_ss_WUs_001g},
from existing \citep{Moser_Kim_Mansour_1999a,
                     Hu_Morfey_Sandham_2006a,
                     Hoyas_Jimenez_2006a,
                     Hoyas_Jimenez_2008a,
                     LozanoDuran_Jimenez_2014a,
                     Bernardini_Pirozzoli_Orlandi_2014a,
                     Vreman_Kuerten_2014b,
                     Vreman_Kuerten_2016a,
                     Lee_Moser_2015a}
\tsn{DNS} computations of turbulent plane channel flow, in the range $Re_{\tau_w}\in[80,5200]$.}
\label{Fig_DTWT_s_DNSDepsij_ss_RetauwIA_sss_AIMWC_001}
\end{figure}

The variation with $Re_{\tau_w}$ of the $y^+$-distributions of $r_{ij}$ and $\varepsilon_{ij}$ anisotropy \figrefsatob{Fig_DTWT_s_DNSDepsij_ss_RetauwIA_sss_ATs_001}
                                                                                                                      {Fig_DTWT_s_DNSDepsij_ss_RetauwIA_sss_AIs_001}
indicates trends in the evolution of turbulence structure with increasing $Re_{\tau_w}$. These trends are better quantified by studying the evolution with $Re_{\tau_w}$ of the anisotropy invariants ($-\II{b}$, $\III{b}$, $-\II{b_\varepsilon}$, $\III{b_\varepsilon}$)
at the wall and at the centerline \figref{Fig_DTWT_s_DNSDepsij_ss_RetauwIA_sss_AIMWC_001}.
Several of the available \tsn{DNS} data \citep{Kim_Moin_Moser_1987a,
                                               Moser_Kim_Mansour_1999a,
                                               Hu_Morfey_Sandham_2002a,
                                               Hu_Morfey_Sandham_2003a,
                                               Hu_Morfey_Sandham_2006a,
                                               delAlamo_Jimenez_2003a,
                                               Hoyas_Jimenez_2006a,
                                               Hoyas_Jimenez_2008a,
                                               LozanoDuran_Jimenez_2014a,
                                               Bernardini_Pirozzoli_Orlandi_2014a,
                                               Vreman_Kuerten_2014a,
                                               Vreman_Kuerten_2014b,
                                               Vreman_Kuerten_2016a,
                                               Lee_Moser_2015a}
acquired since the pioneering work of \citet{Kim_Moin_Moser_1987a} were considered in this study.
These data were obtained by several authors with different computational accuracy indicators (spatio-temporal resolution, box size, observation time and sampling frequency) using
a variety of computational methods and/or codes.
Notice that for the low $Re_{\tau_w}\approxeq180$ case, \citet{Vreman_Kuerten_2014a} recently reported a detailed study demonstrating consistent convergence of \tsn{DNS} results with increasing computational accuracy.
At the centerline, only databases including $\varepsilon_{ij}$-data \citep{Moser_Kim_Mansour_1999a,
                                                                       delAlamo_Jimenez_2003a,
                                                                       Hoyas_Jimenez_2006a,
                                                                       Hoyas_Jimenez_2008a,
                                                                       Vreman_Kuerten_2014b,
                                                                       Vreman_Kuerten_2016a},
were used for the $\tsr{b_\varepsilon}$-invariants \figref{Fig_DTWT_s_DNSDepsij_ss_RetauwIA_sss_AIMWC_001}, and wall-values for the $\tsr{b_\varepsilon}$-invariants were used when available \citep{Moser_Kim_Mansour_1999a,
                                                                                                                                                                                           delAlamo_Jimenez_2003a,
                                                                                                                                                                                           Hoyas_Jimenez_2006a,
                                                                                                                                                                                           Hoyas_Jimenez_2008a}.
For those databases that did not include wall-values \citep{Vreman_Kuerten_2014b,
                                                            Vreman_Kuerten_2016a} or $\varepsilon_{ij}$-data \citep{Hu_Morfey_Sandham_2006a,
                                                                                                                    Bernardini_Pirozzoli_Orlandi_2014a,
                                                                                                                    LozanoDuran_Jimenez_2014a}
wall-invariants were estimated by linear extrapolation of the $\tsr{b}$-invariants from the first 2 grid-points.
The very-near-wall ($y^+\lessapprox0.2$) data of \citet{Bernardini_Pirozzoli_Orlandi_2014a} were noisy
(presumably because of the extreme near-wall cosine-stretching of the wall-normal mesh-size) and were not used;
corresponding wall-values were obtained by extrapolation from the first 2 grid-points with $y^+\gtrapprox0.2$.
The $Re_{\tau_w}\approxeq4180$ small-box data of \citet{LozanoDuran_Jimenez_2014a} were only used at the wall.
Finally, it was found interesting to include the $Re_{\tau_w}<180$ data of \citet{Hu_Morfey_Sandham_2002a,
                                                                                  Hu_Morfey_Sandham_2003a,
                                                                                  Hu_Morfey_Sandham_2006a}
illustrating low-$Re_{\tau_w}$ asymptotics.

It should be stated from the outset that anisotropy invariants are much more sensitive to $r_{ij}$ and $\varepsilon_{ij}$ data uncertainties \citep[p. 5]{Schultz_Flack_2013a}
because of the cumulative propagation of these uncertainties \citep[\S3, pp. 45--91]{Taylor_1997a} in the calculation of the anisotropy tensors \eqrefsab{Eq_DTWT_s_DNSDepsij_ss_A_001a}
                                                                                                                                                         {Eq_DTWT_s_DNSDepsij_ss_A_002a},
and then of their invariants \eqrefsab{Eq_DTWT_s_DNSDepsij_ss_A_001a}
                                      {Eq_DTWT_s_DNSDepsij_ss_A_002a}.
Nonetheless, although data for some of the invariants exhibit substantial scatter \figref{Fig_DTWT_s_DNSDepsij_ss_RetauwIA_sss_AIMWC_001}, it appears that $Re_{\tau_w}$-trends can be deduced with reasonable confidence.

At the wall, the data for both ${\II{b}}_w\stackrel{\eqref{Eq_DTWT_s_DNSDepsij_ss_WA_002c}}{=}{\II{b_{\varepsilon}}}_w$ and ${\III{b}}_w\stackrel{\eqref{Eq_DTWT_s_DNSDepsij_ss_WA_002d}}{=}{\III{b_{\varepsilon}}}_w$ show consistent behaviour with
reasonably small scatter \figref{Fig_DTWT_s_DNSDepsij_ss_RetauwIA_sss_AIMWC_001} suggesting that the near-wall \tsn{DNS} data are quite robust with respect to the variation of computational accuracy indicators of the simulations.
It seems likely that, with increasing $Re_{\tau_w}$, the wall-invariants reach asymptotic values.
On the other hand, as $Re_{\tau_w}$ decreases, both invariants increase sharply \figref{Fig_DTWT_s_DNSDepsij_ss_RetauwIA_sss_AIMWC_001}. Notice that the very-small-box $Re_{\tau_w}\approxeq 4180$ data
of \citet{LozanoDuran_Jimenez_2014a}, which were only considered at the wall,
are consistent with the large-box data \citep{Bernardini_Pirozzoli_Orlandi_2014a,Lee_Moser_2015a} suggesting that the small-box bias mainly impacts the centerline region, whereas the very-near-wall behaviour
is less sensitive to the details of the large-scale outer-flow structures.

At the centerline, the data show considerable scatter \figref{Fig_DTWT_s_DNSDepsij_ss_RetauwIA_sss_AIMWC_001}, especially for the $\tsr{b}$-invariants $(-\II{b},\III{b})_\tsn{CL}$, in line with the observation on the cumulative impact of $r_{ij}$ and $\varepsilon_{ij}$
uncertainties on the calculation of the invariants. Notice first that the scatter is more important for the $\tsr{b}$-invariants which characterize \citep{Lee_Reynolds_1987a} the large-scale anisotropy and are therefore more sensitive to box-size and
observation-time, whereas the $\tsr{b_\varepsilon}$-invariants $(-\II{b_\varepsilon},\III{b_\varepsilon})_\tsn{CL}$  which characterize the small-scales anisotropy are more robust.
With respect to box-size, notice that high-$Re_{\tau_w}$ small-box calculations \citep{LozanoDuran_Jimenez_2014a} overpredict the $\tsr{b}$-invariants at the centerline by a factor 2 (not included in \figrefnp{Fig_DTWT_s_DNSDepsij_ss_RetauwIA_sss_AIMWC_001}).
This is not completely unexpected, and, in view of the consistent wall-invariants obtained in these high-$Re_{\tau_w}$ small-box calculations,
it suggests that the very-large-scale structures influence quite substantially turbulence at the centerline and much less in the very-near-wall (sublayer) region.
The general trend of the $\tsr{b}$-invariants $(-\II{b},\III{b})_\tsn{CL}$ at the centerline \figref{Fig_DTWT_s_DNSDepsij_ss_RetauwIA_sss_AIMWC_001}, is a strong decrease with increasing $Re_{\tau_w}$ in the range $Re_{\tau_w}<400$,
followed by a slight increase, probably reaching an asymptotic state with increasing $Re_{\tau_w}$. Nonetheless there is too much scatter in the data to draw definitive conclusions \figref{Fig_DTWT_s_DNSDepsij_ss_RetauwIA_sss_AIMWC_001}.
Regarding the $\tsr{b_\varepsilon}$-invariants $(-\II{b_\varepsilon},\III{b_\varepsilon})_\tsn{CL}$ at the centerline, there appears a clear trend of asymptotic decrease to $\approxeq0$ as $Re_{\tau_w}$ increases \figref{Fig_DTWT_s_DNSDepsij_ss_RetauwIA_sss_AIMWC_001}.

Observation of $b_{ij}$ at the centerline \figref{Fig_DTWT_s_DNSDepsij_ss_RetauwIA_sss_ATs_001}, where by symmetry $b_{{xy}_\tsn{CL}}=0$, shows that $b_{{yy}_\tsn{CL}} \approxeq b_{{zz}_\tsn{CL}}<b_{{xx}_\tsn{CL}}$ \figref{Fig_DTWT_s_DNSDepsij_ss_RetauwIA_sss_ATs_001}, at least 
for $Re_{\tau_w}\geqslant 180$, implying that the Reynolds-stress tensor at the centerline is axisymmetric rod-like \citep[Fig. 4, p. 3]{Simonsen_Krogstad_2005a}. However, the componentality of $\varepsilon_{ij}$ at the centerline is not as 
obvious \figref{Fig_DTWT_s_DNSDepsij_ss_RetauwIA_sss_ATs_001}, especially as the $\tsr{b_\varepsilon}$-invariants approach 0 at the centerline \figref{Fig_DTWT_s_DNSDepsij_ss_RetauwIA_sss_AIMWC_001}, contrary to the $\tsr{b}$-invariants which seem to reach finite 
asymptotic values at the centerline \figref{Fig_DTWT_s_DNSDepsij_ss_RetauwIA_sss_AIMWC_001}. At the rod-like axisymmetric boundary of the realizability triangle \figref{Fig_DTWT_s_DNSDepsij_ss_RetauwIA_sss_AIMWC_001} 
$\mathrm{II}+3\left(\tfrac{1}{4}\mathrm{III}^2\right )^3=0$, so that $\mathrm{II}^{-1}\left(\mathrm{II}+3\left(\tfrac{1}{4}\mathrm{III}^2\right )^3\right )$ is a diagnostic function whose distance from 0 
denotes departure from rod-like axisymmetric componentality. Contrary to $\tsr{b}_\tsn{CL}$, $\tsr{b_{\varepsilon}}_\tsn{CL}$ at the centerline seems to become increasingly nonaxisymmetric as $Re_{\tau_w}$ increases \figref{Fig_DTWT_s_DNSDepsij_ss_RetauwIA_sss_AIMWC_001}.
Finally, the data of \citet{Hu_Morfey_Sandham_2006a} at very-low $Re_{\tau_w}<180$ seem to indicate departure from rod-like axisymmetry for $\tsr{b}_\tsn{CL}$ \figref{Fig_DTWT_s_DNSDepsij_ss_RetauwIA_sss_AIMWC_001}. At $Re_{\tau_w}\approxeq 180$, the data
of \citet{Kim_Moin_Moser_1987a} also indicate a slight departure from rod-like axisymmetry of $\tsr{b}_\tsn{CL}$, but all other data \citep{Vreman_Kuerten_2016a,Hoyas_Jimenez_2006a,Bernardini_Pirozzoli_Orlandi_2014a,Lee_Moser_2015a}, including the
present calculations, indicate that $\tsr{b}_\tsn{CL}$ is  rod-like axisymmetric at $Re_{\tau_w}\approxeq 180$ \figref{Fig_DTWT_s_DNSDepsij_ss_RetauwIA_sss_AIMWC_001}.
Further \tsn{DNS} at very-low $Re_{\tau_w}< 180$ are therefore needed to verify the departure from rod-like axisymmetry of $\tsr{b}_\tsn{CL}$ with decreasing $Re_{\tau_w}$.
 
%
%
%
%
%
%
%
%
%
\section{$\varepsilon_{ij}$-budgets}\label{DTWT_s_epsijBs}
%
%
%
%
%
%
%
%
%

The dynamics of $\varepsilon_{ij}$ are described by an exact transport equation that can be readily obtained by the fluctuating flow equations \parref{DTWT_s_epsijBs_ss_epsijTEq}.
The budgets of the various terms in the transport equations for $\varepsilon_{ij}$ \eqref{Eq_DTWT_s_epsijBs_ss_epsijTEq_002} are studied for low-Reynolds-number plane channel flow \parref{DTWT_s_epsijBs_ss_epsijTBs}.

Notice first that $\varepsilon_{ij}$ \eqref{Eq_DTWT_s_I_002a} is generated from the 4-order tensor
\begin{subequations}
                                                                                                                                    \label{Eq_DTWT_s_epsijBs_001}
\begin{alignat}{6}
\mathcal{E}_{ijkm}:=2\nu\overline{\dfrac{\partial u'_i}
                                        {\partial  x_k}\dfrac{\partial u'_j}
                                                             {\partial  x_m}}
\stackrel{\eqref{Eq_DTWT_s_I_002a}}{\implies}\varepsilon_{ij}=\mathcal{E}_{ijkm}\delta_{km}=\mathcal{E}_{ijkk}
                                                                                                                                    \label{Eq_DTWT_s_epsijBs_001a}
\end{alignat}
by contraction of the last 2 indices.
Considering $\mathcal{E}_{ijkm}$ is important because this 4-order tensor appears in the production mechanisms of $\varepsilon_{ij}$ \eqref{Eq_DTWT_s_epsijBs_ss_epsijTEq_002}.
The tensor $\mathcal{E}_{ijkm}$ was used by other authors \citep{Durbin_Speziale_1991a,
                                                                 Speziale_Gatski_1997a} 
who studied  $\varepsilon_{ij}$-transport in homogeneous turbulence.
It is also interesting to note that the 9 transport equations for the variances of the fluctuating velocity-gradients studied by \citet[(7, 9), p. 4]{Vreman_Kuerten_2014b}
are also included in the transport equations for $\mathcal{E}_{ijkm}$ \eqref{Eq_DTWT_s_epsijBs_001a} because
\begin{alignat}{6}
2\nu\overline{\left(\dfrac{\partial u_i'}{\partial x_j}\right)^2}\in\Big\{\mathcal{E}_{xxxx},\mathcal{E}_{xxyy},\mathcal{E}_{xxzz},
                                                                          \mathcal{E}_{yyxx},\mathcal{E}_{yyyy},\mathcal{E}_{yyzz},
                                                                          \mathcal{E}_{zzxx},\mathcal{E}_{zzyy},\mathcal{E}_{zzzz}\Big\}
                                                                                                                                    \label{Eq_DTWT_s_epsijBs_001b}
\end{alignat}
\end{subequations}

%
%
%
%
%
\subsection{$\varepsilon_{ij}$-transport equation}\label{DTWT_s_epsijBs_ss_epsijTEq}
%
%
%
%
%

Starting from the fluctuating continuity \citep[(4.6), p. 76]{Mathieu_Scott_2000a}
\begin{subequations}
                                                                                                                                    \label{Eq_DTWT_s_epsijBs_ss_epsijTEq_001}
\begin{alignat}{6}
                         \dfrac{\partial    u'_\ell}
                               {\partial     x_\ell}=0
                                                                                                                                    \label{Eq_DTWT_s_epsijBs_ss_epsijTEq_001a}
\end{alignat}
and fluctuating momentum \citep[(4.31), p. 85]{Mathieu_Scott_2000a}
\begin{alignat}{6}
\rho\dfrac{\partial u'_i}
          {\partial    t}+\rho\bar u_\ell\dfrac{\partial   u'_i}
                                               {\partial x_\ell} =&-\rho\dfrac{\partial       }
                                                                              {\partial x_\ell}\left(u'_iu'_\ell-\overline{u'_iu'_\ell}\right)
                                                                   -\rho u'_\ell\dfrac{\partial\bar u_i}
                                                                                      {\partial  x_\ell}-\dfrac{\partial  p'}
                                                                                                               {\partial x_i}+\mu\dfrac{\partial^2                u'_i}
                                                                                                                                       {\partial x_\ell\partial x_\ell}
                                                                                                                                    \label{Eq_DTWT_s_epsijBs_ss_epsijTEq_001b}
\end{alignat}
\end{subequations}
equations we can work out the transport equations for $\mathcal{E}_{ijkm}$ \eqref{Eq_DTWT_s_epsijBs_001a}, and by contraction \eqref{Eq_DTWT_s_epsijBs_001a} the
transport equation for $\varepsilon_{ij}$, which reads
\begin{alignat}{6}
&\underbrace{\rho\dfrac{\partial\varepsilon_{ij}}
                       {\partial               t}+\rho\bar u_\ell\dfrac{\partial\varepsilon_{ij}}
                                                                       {\partial          x_\ell}}_{\displaystyle C_{\varepsilon_{ij}}}=
  \underbrace{\dfrac{\partial       }
                    {\partial x_\ell}\left[\mu\dfrac{\partial\varepsilon_{ij}}
                                                    {\partial x_\ell          }\right]}_{\displaystyle d_{\varepsilon_{ij}}^{(\mu)}}
 +\underbrace{\dfrac{\partial       }
                    {\partial x_\ell}\left[-\rho\left(\overline{u'_\ell2\nu\dfrac{\partial u'_i}
                                                                                 {\partial  x_k}\dfrac{\partial u'_j}
                                                                                                      {\partial  x_k}}\right)\right]}_{\displaystyle d_{\varepsilon_{ij}}^{(u)}}
                                                                                                                                    \notag\\
&\underbrace{-\rho\varepsilon_{i\ell}\dfrac{\partial\bar u_j}
                                           {\partial  x_\ell}
             -\rho\varepsilon_{j\ell}\dfrac{\partial\bar u_i}
                                           {\partial  x_\ell}}_{\displaystyle P_{\varepsilon_{ij}}^{(1)}}
 \underbrace{-\rho\left(2\nu\overline{\dfrac{\partial u'_i}
                                            {\partial  x_k}\dfrac{\partial   u'_j}
                                                                 {\partial x_\ell}}\right)\left(\dfrac{\partial\bar u_k   }
                                                                                                      {\partial     x_\ell}+\dfrac{\partial\bar u_\ell}
                                                                                                                                  {\partial     x_k   }\right)}_{\displaystyle P_{\varepsilon_{ij}}^{(2)}}
                                                                                                                                    \notag\\
&\underbrace{-\rho\left(2\nu\overline{u'_\ell\dfrac{\partial u'_i}
                                                   {\partial  x_k}}\right)\dfrac{\partial^2 \bar u_j}
                                                                                {\partial x_\ell \partial x_k}
             -\rho\left(2\nu\overline{u'_\ell\dfrac{\partial u'_j}
                                                   {\partial  x_k}}\right)\dfrac{\partial^2 \bar u_i}
                                                                                {\partial x_\ell \partial x_k}}_{\displaystyle P_{\varepsilon_{ij}}^{(3)}}
 \underbrace{-\rho\left[2\nu\overline{\dfrac{\partial u'_\ell}
                                            {\partial     x_k}\left(\dfrac{\partial u'_i}
                                                                          {\partial  x_k}\dfrac{\partial   u'_j}
                                                                                               {\partial x_\ell}
                                                                   +\dfrac{\partial u'_j}
                                                                          {\partial  x_k}\dfrac{\partial   u'_i}
                                                                                               {\partial x_\ell}\right)}\right]}_{\displaystyle \Xi_{\varepsilon_{ij}}=:P_{\varepsilon_{ij}}^{(4)}}
                                                                                                                                    \notag\\
&\underbrace{-2\nu\overline{\dfrac{\partial u'_i}
                                  {\partial  x_k}\dfrac{\partial^2            p'}
                                                       {\partial x_j\partial x_k}}
             -2\nu\overline{\dfrac{\partial u'_j}
                                  {\partial  x_k}\dfrac{\partial^2            p'}
                                                       {\partial x_i\partial x_k}}}_{\displaystyle \Pi_{\varepsilon_{ij}}}
 -\underbrace{\rho\overline{\left(2\nu\dfrac{\partial^2 u'_i}
                                            {\partial  x_k\partial x_\ell}\right)\left(2\nu\dfrac{\partial^2 u'_j}
                                                                                                 {\partial  x_k\partial x_\ell}\right)}}_{\displaystyle \rho\varepsilon_{\varepsilon_{ij}}}
                                                                                                                                    \label{Eq_DTWT_s_epsijBs_ss_epsijTEq_002}
\end{alignat}
In \eqref{Eq_DTWT_s_epsijBs_ss_epsijTEq_002}
$C_{\varepsilon_{\ij}}$ is the convection of $\varepsilon_{\ij}$ by the mean-flow velocity field $\bar u_\ell$,
$d_{\varepsilon_{ij}}^{(\mu)}$ is the diffusion of $\varepsilon_{\ij}$ by molecular viscosity (notice that $d_{\varepsilon_{ij}}^{(\mu)}\stackrel{\eqref{Eq_DTWT_s_epsijBs_ss_epsijTEq_002}}{=}\mu\nabla^2\varepsilon_{ij}$ for $\mu=\const$),
$d_{\varepsilon_{ij}}^{(u)}$ is the turbulent diffusion (mixing) of $\varepsilon_{\ij}$ by the fluctuating velocity field $u_\ell'$,
$P_{\varepsilon_{ij}}^{(1)}$ is the production of $\varepsilon_{\ij}$ by the action of its components on the mean-flow velocity-gradients,
$P_{\varepsilon_{ij}}^{(2)}$ is the production of $\varepsilon_{\ij}$ by the action of $\mathcal{E}_{ijkm}$ \eqref{Eq_DTWT_s_epsijBs_001} on the mean-flow velocity-gradients,
$P_{\varepsilon_{ij}}^{(3)}$ is the production of $\varepsilon_{\ij}$ by the mean-flow velocity-Hessian,
$P_{\varepsilon_{ij}}^{(4)}:=\Xi_{\varepsilon_{ij}}$ corresponds to triple correlations of fluctuating velocity-gradients whose trace $\tfrac{1}{2}\Xi_{\varepsilon_{\ell\ell}}$ was identified by
                                                     \citet{Mansour_Kim_Moin_1988a} as a term "producing" (gain in the budgets of) $\varepsilon:=\tfrac{1}{2}\varepsilon_{\ell\ell}$
                                                     (by extension the denomination $P_{\varepsilon_{ij}}^{(4)}$ is used, although this term does not contain gradients of the mean-flow field),
$\Pi_{\varepsilon_{ij}}$ are the terms containing the fluctuating pressure-Hessian, and
$\varepsilon_{\varepsilon_{ij}}$ is the destruction of $\varepsilon_{\ij}$ by the action of molecular viscosity.
Following usual practice \citep[(1), p. 17]{Mansour_Kim_Moin_1988a} for the incompressible $r_{ij}$ equations \eqref{Eq_DTWT_s_I_001} and for the $\varepsilon$ equation \citep[(23), p. 23]{Mansour_Kim_Moin_1988a},
a computable viscous diffusion term $d_{\varepsilon_{ij}}^{(\mu)}$ was chosen in \eqref{Eq_DTWT_s_epsijBs_ss_epsijTEq_002},
with a corresponding appropriate definition of the destruction-of-dissipation tensor $\varepsilon_{\varepsilon_{ij}}$ \eqref{Eq_DTWT_s_epsijBs_ss_epsijTEq_002},
in lieu of alternative splittings based on the viscous-stress tensor \citep[(2), p. 189]{BenNasr_Gerolymos_Vallet_2014a}.

Using \eqref{Eq_DTWT_s_epsijBs_001} the production of $\varepsilon_{ij}$ by mean-flow velocity-gradients reads
\begin{alignat}{6}
P_{\varepsilon_{ij}}^{(1)}+P_{\varepsilon_{ij}}^{(2)}\stackrel{\eqrefsab{Eq_DTWT_s_epsijBs_ss_epsijTEq_002}
                                                                        {Eq_DTWT_s_epsijBs_001a}}{=}
 \underbrace{-\rho\mathcal{E}_{i\ell kk}\dfrac{\partial\bar u_j}
                                              {\partial  x_\ell}
             -\rho\mathcal{E}_{j\ell kk}\dfrac{\partial\bar u_i}
                                              {\partial  x_\ell}}_{\displaystyle P_{\varepsilon_{ij}}^{(1)}}
 \underbrace{-\rho\mathcal{E}_{ijk\ell}\left(\dfrac{\partial\bar u_k   }
                                                    {\partial     x_\ell}+\dfrac{\partial\bar u_\ell}
                                                                                {\partial     x_k   }\right)}_{\displaystyle P_{\varepsilon_{ij}}^{(2)}}
                                                                                                                                    \label{Eq_DTWT_s_epsijBs_ss_epsijTEq_004}
\end{alignat}
highlighting the importance of considering the generating 4-order tensor $\mathcal{E}_{ijkm}$ \eqref{Eq_DTWT_s_epsijBs_001}. The separation in 2 terms,
$P_{\varepsilon_{ij}}^{(1)}$ and $P_{\varepsilon_{ij}}^{(2)}$, was made to distinguish between the computable (from the knowledge of $\varepsilon_{ij}$ and the mean-flow field) term $P_{\varepsilon_{ij}}^{(1)}$,
and $P_{\varepsilon_{ij}}^{(2)}$ which involves components of $\mathcal{E}_{ijkm}$ that do not simplify by contraction \eqref{Eq_DTWT_s_epsijBs_001a} to $\varepsilon_{ij}$.

%
%
%
%
%
\subsection{$\varepsilon_{ij}$-transport budgets}\label{DTWT_s_epsijBs_ss_epsijTBs}
%
%
%
%
%

\begin{figure}
\begin{center}
\begin{picture}(450,370)
\put(0,0){\includegraphics[angle=0,width=380pt,bb=59 280 516 721]{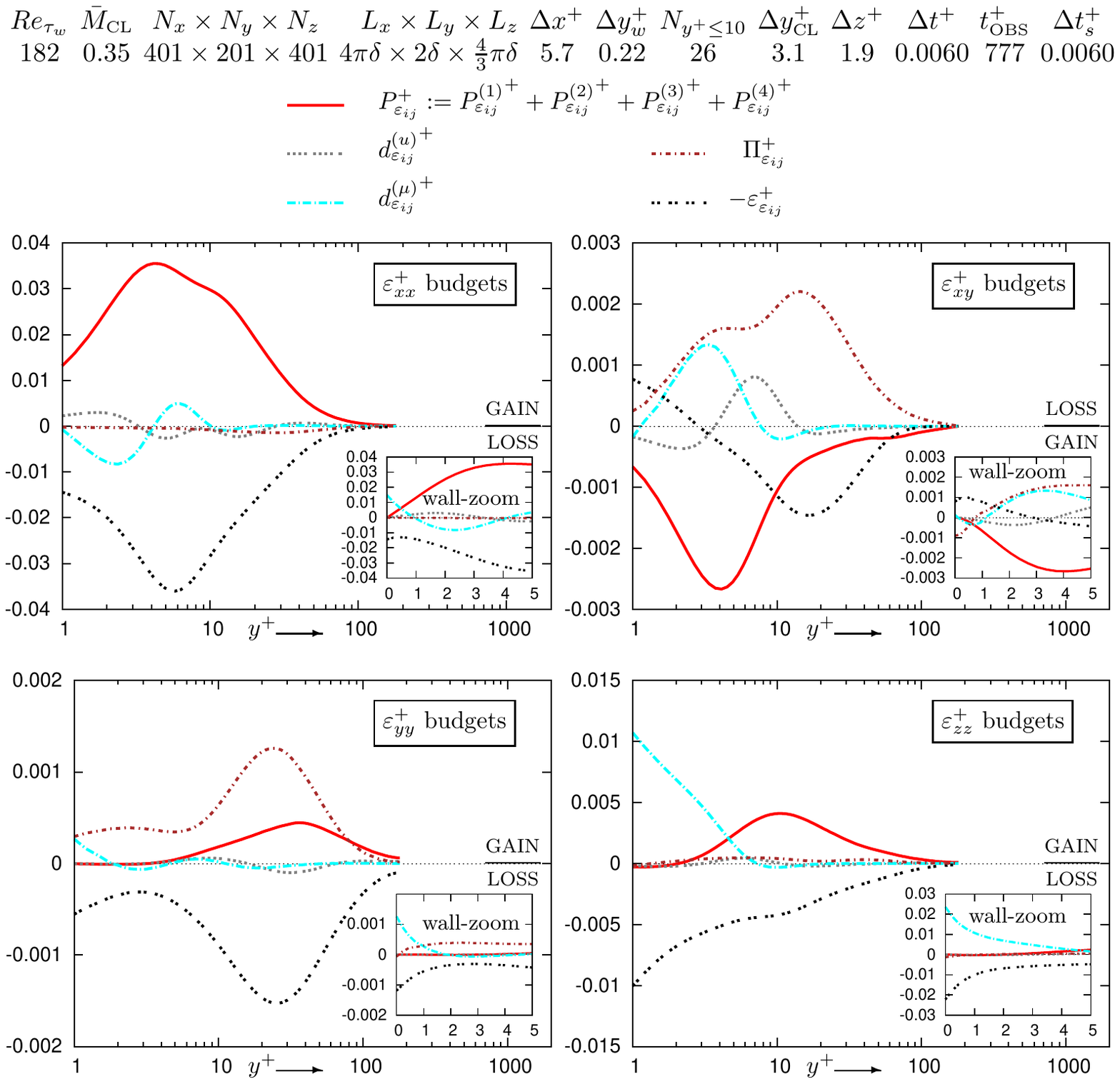}}
\end{picture}
\end{center}
\caption{Budgets, in wall-units \eqref{Eq_DTWT_s_AppendixABVSy+0_ss_WUs_001f}, of the transport equations \eqref{Eq_DTWT_s_epsijBs_ss_epsijTEq_002}
for the dissipation tensor $\varepsilon_{ij}$ \eqref{Eq_DTWT_s_I_002a}, from the present \tsn{DNS} computations of turbulent plane channel flow ($Re_{\tau_w}\approxeq182$, $\bar M_\tsn{CL}\approxeq0.35$),
plotted against the inner-scaled \eqref{Eq_DTWT_s_AppendixABVSy+0_ss_WUs_001c} wall-distance $y^+$ (logscale and linear wall-zoom).}
\label{Fig_DTWT_s_epsijBs_ss_epsijTBs_001}
\end{figure}

Budgets of the $\varepsilon_{ij}$-transport equations \eqref{Eq_DTWT_s_epsijBs_ss_epsijTEq_002}, for turbulent plane channel flow, were obtained \figref{Fig_DTWT_s_epsijBs_ss_epsijTBs_001}
in an $L_x\times L_y\times L_z=4\pi\delta\times2\delta\times\tfrac{4}{3}\pi\delta$ computational box, using a carefully validated \tsn{DNS} solver \citep{Gerolymos_Senechal_Vallet_2010a,Gerolymos_Senechal_Vallet_2013a,Gerolymos_Vallet_2014a}.
The resolution in the homogeneous $xz$-directions is $\Delta x^+\approxeq 5.7$ and $\Delta z^+\approxeq 1.9$. The wall-normal size of the grid cells adjacent to the wall was $\Delta y_w^+\approxeq 0.22$. More importantly, the mesh 
(65\% of the nodes were stretched near the walls geometrically  with ratio $r_j=1.0427$, the remaining nodes in the outer region being equidistant) was kept fine in the entire near-wall region, with 
$N_{y^+\leq 10}=26$ grid cells between the wall and $y^+\approxeq 10$, and remained fine up to the centerline where the wall-normal cell-size was $\Delta y_\tsn{CL}^+\approxeq 3.1$. This spatial resolution, combined with
the $O(\Delta\ell^{17})$ scheme used \citep{Gerolymos_Senechal_Vallet_2009a,Gerolymos_Senechal_Vallet_2010a} is quite fine in view of current state-of-the-art \tsn{DNS} at this $Re_{\tau_w}\approxeq 180$ \citep{Vreman_Kuerten_2014a,Vreman_Kuerten_2016a}.
Statistics were acquired at high sampling frequency (at every iteration; $\Delta t_s^+=\Delta t^+\approxeq 0.0060$, albeit for a relatively short observation time $t_{\tsn{OBS}}^+\approxeq 777$). The close agreement \parref{DTWT_s_RetauwIoepsijBs}
of the present results for the budgets of the diagonal components with the highly resolved computations of \citet{Vreman_Kuerten_2016a} further substantiate the validity of the computations.
Wall-asymptotics of the various terms in \eqref{Eq_DTWT_s_epsijBs_ss_epsijTEq_002} can be obtained using the Taylor-expansions \eqref{Eq_DTWT_s_DNSDepsij_ss_WA_001}
in the fluctuating-momentum equations \eqref{Eq_DTWT_s_epsijBs_ss_epsijTEq_001b}. Although a full report of these calculations is outside the scope of this paper,
some of the limiting wall-values obtained from this procedure are included in the following discussion.

Regarding the streamwise component $\varepsilon_{xx}^+$ \figref{Fig_DTWT_s_epsijBs_ss_epsijTBs_001}, production $P_{\varepsilon_{xx}}^+$ (gain) roughly balances destruction $-\varepsilon_{\varepsilon_{xx}}^+$ (loss) in the major part of the channel ($y^+\gtrapprox 1$),
with lesser contributions of diffusion ($d_{\varepsilon_{xx}}^{(u)+}+d_{\varepsilon_{xx}}^{(\mu)+}$). Production $P_{\varepsilon_{xx}}^+$  peaks at $y^+\approxeq 4$ and destruction $\varepsilon_{\varepsilon_{xx}}^+$ at $y^+\approxeq 5$.
The pressure term $\Pi_{\varepsilon_{xx}}^+$ is negligible in the budgets of $\varepsilon_{xx}^+$ throughout the channel. Very near the wall ($y^+\lessapprox1$; \figrefnp{Fig_DTWT_s_epsijBs_ss_epsijTBs_001})  
production \smash{$P_{\varepsilon_{xx}}^+\underset{y^+\to 0}{\to 0}$} (wall-asymptotic expansion; \parrefnp{DTWT_s_DNSDepsij_ss_WA})
and viscous diffusion $d_{\varepsilon_{xx}}^{(\mu)+}$ (gain $\forall\;y^+\lessapprox1$) roughly counters destruction $-\varepsilon_{\varepsilon_{xx}}^+$ ($y^+\to 0$; \figrefnp{Fig_DTWT_s_epsijBs_ss_epsijTBs_001}).
At the wall, the pressure term $[\Pi_{\varepsilon_{xx}}^+]_w=8\overline{B_v'\partial_xA_u'}^+\neq 0$ (wall-asymptotic expansion; \parrefnp{DTWT_s_DNSDepsij_ss_WA}).
However $\abs{\Pi_{\varepsilon_{xx}}^+}_w\ll [\varepsilon_{\varepsilon_{xx}}^+]_w \approxeq [d_{\varepsilon_{xx}}^{(\mu)+}]_w$ \figref{Fig_DTWT_s_epsijBs_ss_epsijTBs_001}.
%
\begin{figure}
\begin{center}
\begin{picture}(450,370)
\put(0,0){\includegraphics[angle=0,width=380pt,bb=59 280 516 721]{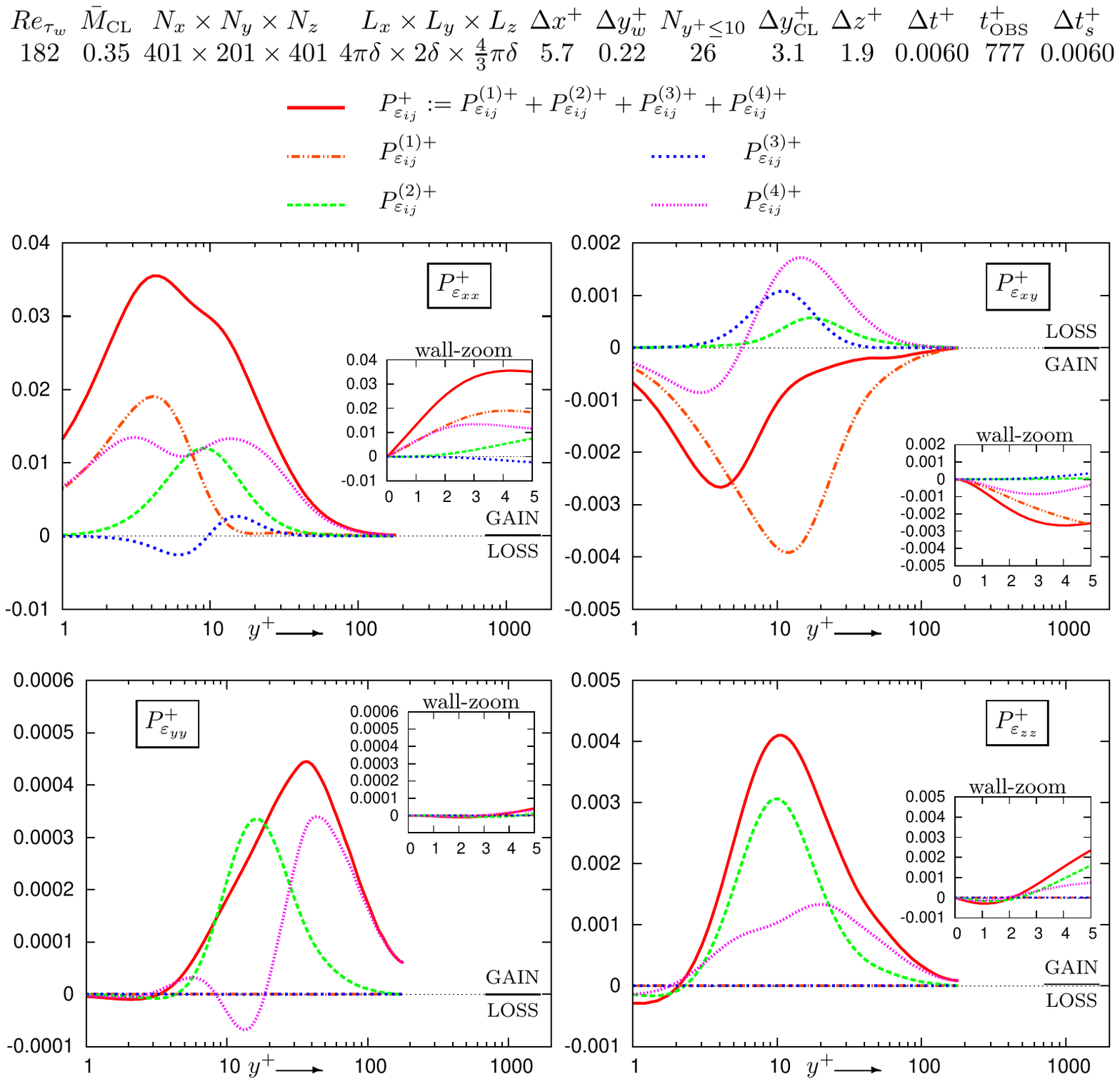}}
\end{picture}
\end{center}
\caption{Comparison of the various mechanisms of production $P_{\varepsilon_{ij}}=P^{(1)}_{\varepsilon_{ij}}+P^{(2)}_{\varepsilon_{ij}}+P^{(3)}_{\varepsilon_{ij}}+P^{(4)}_{\varepsilon_{ij}}$ \eqref{Eq_DTWT_s_epsijBs_ss_epsijTEq_002}
of the dissipation tensor $\varepsilon_{ij}$ \eqref{Eq_DTWT_s_I_002a}, in wall-units \eqref{Eq_DTWT_s_AppendixABVSy+0_ss_WUs_001f},
from the present \tsn{DNS} computations of turbulent plane channel flow ($Re_{\tau_w}\approxeq182$, $\bar M_\tsn{CL}\approxeq0.35$),
plotted against the inner-scaled \eqref{Eq_DTWT_s_AppendixABVSy+0_ss_WUs_001c} wall-distance $y^+$ (logscale and linear wall-zoom).}
\label{Fig_DTWT_s_epsijBs_ss_epsijTBs_002}
\end{figure}
%
\begin{figure}
\begin{center}
\begin{picture}(450,320)
\put(0,0){\includegraphics[angle=0,width=380pt,bb=60 280 516 662]{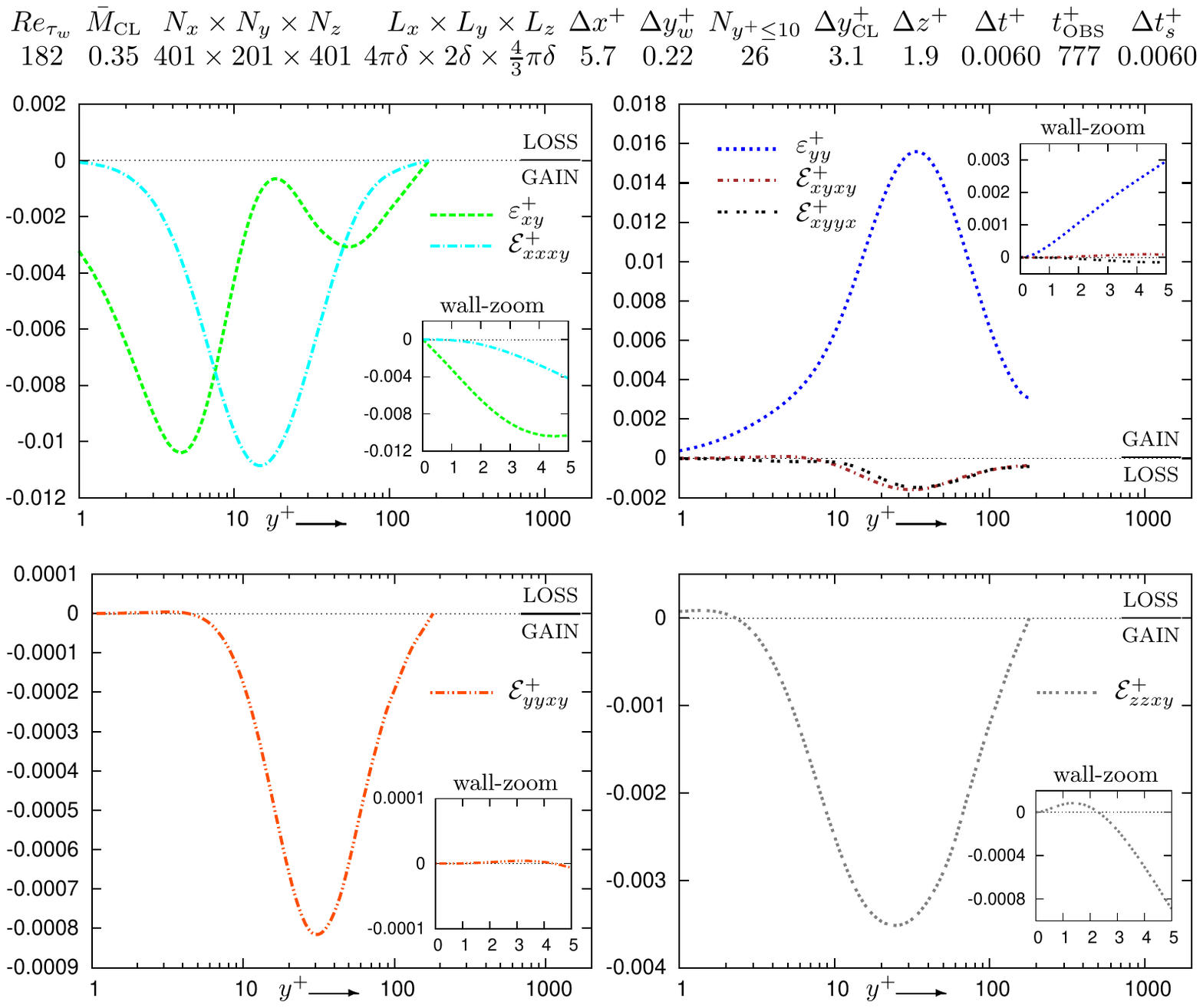}}
\end{picture}
\end{center}
\caption{Components of $\varepsilon_{ij}$ \eqref{Eq_DTWT_s_I_002a} and of $\mathcal{E}_{ijkm}$ \eqref{Eq_DTWT_s_epsijBs_001} that appear in the production by mean-velocity-gradient
$P^{(1)}_{\varepsilon_{ij}}$ and $P^{(2)}_{\varepsilon_{ij}}$ terms \eqref{Eq_DTWT_s_epsijBs_ss_epsijTEq_002}
of the $\varepsilon_{ij}$-transport budgets expressed for the particular case of fully developed plane channel flow \tabref{Tab_DTWT_s_epsijBs_ss_epsijTBs_001},
in wall-units \eqref{Eq_DTWT_s_AppendixABVSy+0_ss_WUs_001e},
from the present \tsn{DNS} computations of turbulent plane channel flow ($Re_{\tau_w}\approxeq182$, $\bar M_\tsn{CL}\approxeq0.35$),
plotted against the inner-scaled \eqref{Eq_DTWT_s_AppendixABVSy+0_ss_WUs_001c} wall-distance $y^+$ (logscale and linear wall-zoom).}
\label{Fig_DTWT_s_epsijBs_ss_epsijTBs_003}
\end{figure}

The budgets of the spanwise component $\varepsilon_{zz}^+$ \figref{Fig_DTWT_s_epsijBs_ss_epsijTBs_001} are also dominated by a balance between production $P_{\varepsilon_{zz}}^+$ (gain) and destruction $-\varepsilon_{\varepsilon_{zz}}^+$ (loss) in
the major part of the channel ($y^+\gtrapprox 5$). However, production $P_{\varepsilon_{zz}}^+$ peaks at $y^+\approxeq 10$ and becomes negligible at $y^+\approxeq 3$
as \smash{$P_{\varepsilon_{zz}}^+\underset{y^+\to 0}{\to 0}$} (wall-asymptotic expansion; \parrefnp{DTWT_s_DNSDepsij_ss_WA}).
In the near-wall region, again destruction $-\varepsilon_{\varepsilon_{zz}}^+$ (loss $\forall\;y^+$) is balanced by viscous diffusion $d_{\varepsilon_{zz}}^{(\mu)+}$ (gain $\forall\;y^+\lessapprox6$),
but this zone extends further away from the wall ($y^+\lessapprox 3$; \figrefnp{Fig_DTWT_s_epsijBs_ss_epsijTBs_001}) compared
to the streamwise component ($y^+\lessapprox1$; \figrefnp{Fig_DTWT_s_epsijBs_ss_epsijTBs_001}). As for the streamwise component,
the other mechanisms, $d_{\varepsilon_{zz}}^{(u)+}$ and $\Pi_{\varepsilon_{zz}}^+$, have a very small contribution to the $\varepsilon_{zz}^+$-budget. Another difference between the streamwise and the spanwise
components is that $\varepsilon_{\varepsilon_{zz}}^+$ is maximum at the wall decreasing monotonically with $y^+$, contrary to $\varepsilon_{\varepsilon_{xx}}^+$ \figref{Fig_DTWT_s_epsijBs_ss_epsijTBs_001}.
\begin{table}
\begin{center}
\scalefont{1.}{
\begin{alignat}{6}
P_{\varepsilon_{xx}}^{(1)}&=&-2\rho\varepsilon_{xy}\dfrac{d\bar u}
                                                       {dy     }
\qquad\qquad
P_{\varepsilon_{xx}}^{(2)}&=&-2\rho\mathcal{E}_{xxxy}\dfrac{d\bar u}
                                                         {dy     }
\qquad\qquad
P_{\varepsilon_{xx}}^{(3)}&=&-4\rho\nu\overline{v'\dfrac{\partial u'}
                                                      {\partial  y}}\dfrac{d^2\bar u}
                                                                          {dy^2     }
                                                                                                                                    \notag\\
P_{\varepsilon_{xy}}^{(1)}&=&-\rho\varepsilon_{yy}\dfrac{d\bar u}
                                                      {dy     }
\qquad\qquad
P_{\varepsilon_{xy}}^{(2)}&=&-\rho\left(\mathcal{E}_{xyxy}+\mathcal{E}_{xyyx}\right)\dfrac{d\bar u}
                                                                                       {dy     }
\qquad\qquad
P_{\varepsilon_{xy}}^{(3)}&=&-\rho2\nu\overline{v'\dfrac{\partial v'}
                                                      {\partial  y}}\dfrac{d^2\bar u}
                                                                          {dy^2     }
                                                                                                                                    \notag\\
P_{\varepsilon_{yy}}^{(1)}&=&0
\qquad\qquad
P_{\varepsilon_{yy}}^{(2)}&=&-2\rho\mathcal{E}_{yyxy}\dfrac{d\bar u}
                                                         {dy     }
\qquad\qquad
P_{\varepsilon_{yy}}^{(3)}&=&0
                                                                                                                                    \notag\\
P_{\varepsilon_{zz}}^{(1)}&=&0
\qquad\qquad
P_{\varepsilon_{zz}}^{(2)}&=&-2\rho\mathcal{E}_{zzxy}\dfrac{d\bar u}
                                                         {dy     }
\qquad\qquad
P_{\varepsilon_{zz}}^{(3)}&=&0
                                                                                                                                    \notag
\end{alignat}
}

\caption{Components of different mechanisms of production $P_{\varepsilon_{ij}}=P^{(1)}_{\varepsilon_{ij}}+P^{(2)}_{\varepsilon_{ij}}+P^{(3)}_{\varepsilon_{ij}}+P^{(4)}_{\varepsilon_{ij}}$ \eqref{Eq_DTWT_s_epsijBs_ss_epsijTEq_002}
         for fully developed turbulent plane channel.}
\label{Tab_DTWT_s_epsijBs_ss_epsijTBs_001}
\end{center}
\end{table}

The major difference in the budgets of the wall-normal component $\varepsilon_{yy}^+$, compared to the other two diagonal components, is the importance of the pressure term $\Pi_{\varepsilon_{yy}}^+$ which is the dominant gain mechanism throughout the
channel \figref{Fig_DTWT_s_epsijBs_ss_epsijTBs_001}, except very near the wall ($y^+\lessapprox \tfrac{1}{2}$; \figrefnp{Fig_DTWT_s_epsijBs_ss_epsijTBs_001})
where $\Pi_{\varepsilon_{yy}}^+\underset{y^+\to 0}{\to 0}$ (wall-asymptotic expansion; \parrefnp{DTWT_s_DNSDepsij_ss_WA}) and, as for $\varepsilon_{xx}^+$ and $\varepsilon_{zz}^+$,
viscous diffusion $d_{\varepsilon_{yy}}^{(\mu)+}$ (gain) balances destruction $-\varepsilon_{\varepsilon_{yy}}^+$ ($y^+\lessapprox \tfrac{1}{2}$; \figrefnp{Fig_DTWT_s_epsijBs_ss_epsijTBs_001}).
Production $P_{\varepsilon_{yy}}^+$ (gain) is significant in the buffer region ($10\lessapprox y^+ \lessapprox 100$; \figrefnp{Fig_DTWT_s_epsijBs_ss_epsijTBs_001}) although it becomes comparable to $\Pi_{\varepsilon_{yy}}^+$ only for $y^+\gtrapprox100$
\figref{Fig_DTWT_s_epsijBs_ss_epsijTBs_001}.
Turbulent diffusion $d_{\varepsilon_{yy}}^{(u)+}$ is generally weak throughout the channel \figref{Fig_DTWT_s_epsijBs_ss_epsijTBs_001}. 

The budgets of the shear component $\varepsilon_{xy}^+<0\;\forall\,y^+>0$ \citep[Fig. 4, p. 19]{Mansour_Kim_Moin_1988a} exhibit a fundamentally different behaviour compared to the diagonal components \figref{Fig_DTWT_s_epsijBs_ss_epsijTBs_001}. All of the terms in the
$\varepsilon_{xy}^+$-budgets ($P_{\varepsilon_{xy}}^+$, $\Pi_{\varepsilon_{xy}}^+$, $d_{\varepsilon_{xy}}^{(u)+}$, $d_{\varepsilon_{xy}}^{(\mu)+}$, $-\varepsilon_{\varepsilon_{xy}}^+$) are significant \figref{Fig_DTWT_s_epsijBs_ss_epsijTBs_001}.
Contrary to the diagonal components, the destruction term $-\varepsilon_{\varepsilon_{xy}}^+$ contributes as gain to the budgets in the major part of the channel ($y^+\gtrapprox 3$; \figrefnp{Fig_DTWT_s_epsijBs_ss_epsijTBs_001}),
becoming  an actual destruction mechanism (loss) only in the viscous sublayer ($y^+\lessapprox3$; \figrefnp{Fig_DTWT_s_epsijBs_ss_epsijTBs_001}).
Notice that, at the wall, viscous diffusion of the shear component is substantially weaker than the 2 other mechanisms present in \eqref{Eq_DTWT_s_epsijBs_ss_epsijTEq_002}, \viz\
$\abs{\Pi_{\varepsilon_{xy}}^+}_w\gg\abs{d_{\varepsilon_{xy}}^{(\mu)+}}_w=\abs{\varepsilon_{\varepsilon_{xy}}^+-\Pi_{\varepsilon_{xy}}^+}_w\ll\abs{\varepsilon_{\varepsilon_{xy}}^+}_w$
(wall-asymptotic expansion; \parrefnp{DTWT_s_DNSDepsij_ss_WA}; \figrefnp{Fig_DTWT_s_epsijBs_ss_epsijTBs_001}).
In a large part of the channel, $-\varepsilon_{\varepsilon_{xy}}^+<0\,\forall\,y^+\gtrapprox3$ \figref{Fig_DTWT_s_epsijBs_ss_epsijTBs_001} is an important gain mechanism, along 
with production $P_{\varepsilon_{xy}}^+<0\,\forall\,y^+>0$ \figref{Fig_DTWT_s_epsijBs_ss_epsijTBs_001}.
On the other hand, $\Pi_{\varepsilon_{xy}}^+$ is the main loss mechanism in the major part of the channel \figref{Fig_DTWT_s_epsijBs_ss_epsijTBs_001}.
Notice the complicated $y^+$-evolution of $\Pi_{\varepsilon_{xy}}^+$ \figref{Fig_DTWT_s_epsijBs_ss_epsijTBs_001},
which is $<0$ (gain) at the wall, crossing to $>0$ (loss) at $y^+\gtrapprox\tfrac{1}{2}$, and presents a plateau ($3\lessapprox y^+\lessapprox7$) followed by a global maximum at $y^+\approxeq15$, before decreasing monotonically to $0$ at the centerline.
In the region $1\lessapprox y^+\lessapprox20$, both diffusion mechanisms, $d_{\varepsilon_{xy}}^{(\mu)+}$ and $d_{\varepsilon_{xy}}^{(u)+}$, contribute significantly to the $\varepsilon_{xy}$-budgets \figref{Fig_DTWT_s_epsijBs_ss_epsijTBs_001}.

A major difference between the $r_{ij}$-budgets \citep{Mansour_Kim_Moin_1988a} and the $\varepsilon_{ij}$-budgets \figref{Fig_DTWT_s_epsijBs_ss_epsijTBs_001} lies in the production mechanisms  $P_{ij}$ \eqref{Eq_DTWT_s_I_001}
and $P_{\varepsilon_{ij}}$ \eqref{Eq_DTWT_s_epsijBs_ss_epsijTEq_002}. In the $r_{ij}$-budgets of plane channel flow, only $P_{xx}^+\neq0\neq P_{xy}^+$ whereas $P_{yy}^+=P_{zz}^+=0$ $\forall\;y^+$ \citep[Figs. 2 and 3, pp. 18--19]{Mansour_Kim_Moin_1988a}. 
On the contrary, all of the components 
$P_{\varepsilon_{ij}}\neq 0$ $\forall\;y^+>0$ in general \figref{Fig_DTWT_s_epsijBs_ss_epsijTBs_001}.
In relation to the above observation, the main gain mechanism in the $\varepsilon_{zz}^+$-budgets is production $P_{\varepsilon_{zz}}^+$ \figref{Fig_DTWT_s_epsijBs_ss_epsijTBs_001}
in contrast to the $r_{zz}$-budgets \citep[Fig. 3, p. 19]{Mansour_Kim_Moin_1988a} where, in the absence of production $P_{zz}^+=0\;\forall\;y^+$, the pressure term $\Pi_{zz}^+$ is the main gain mechanism.
On the contrary, $\Pi_{\varepsilon_{zz}}^+$ has negligible contribution to the $\varepsilon_{zz}$-budgets \figref{Fig_DTWT_s_epsijBs_ss_epsijTBs_001}.
It is noteworthy that the importance of the pressure terms $\Pi_{\varepsilon_{yy}}^+$ and $\Pi_{\varepsilon_{xy}}^+$ in the budgets of the $\varepsilon_{yy}^+$ and $\varepsilon_{xy}^+$ components \figref{Fig_DTWT_s_epsijBs_ss_epsijTBs_001} is also quite 
generally observed in the budgets of wall-normal fluxes, including the Reynolds stresses $r_{yy}^+$ \citep[Fig. 2, p. 18]{Mansour_Kim_Moin_1988a} and $r_{xy}^+$ \citep[Fig. 4, p. 19]{Mansour_Kim_Moin_1988a}, and in the compressible case, 
the fluxes of thermodynamic quantities such as temperature $\overline{T'v'}$ \citep[Figs.~14 and 15, pp.~737--738]{Gerolymos_Vallet_2014a}, density $\overline{\rho'v'}$ \citep[Fig.~12, p. 731]{Gerolymos_Vallet_2014a},
pressure $\overline{p'v'}$ \citep[Fig.~16, p. 741]{Gerolymos_Vallet_2014a} and entropy $\overline{s'v'}$ \citep[Fig.~13, p. 734]{Gerolymos_Vallet_2014a}.

Production $P_{\varepsilon_{ij}}^+$ \figref{Fig_DTWT_s_epsijBs_ss_epsijTBs_001} of $\varepsilon_{ij}$ contains 4 different mechanisms,
$P_{\varepsilon_{ij}}^+=P_{\varepsilon_{ij}}^{(1)+}+P_{\varepsilon_{ij}}^{(2)+}+P_{\varepsilon_{ij}}^{(3)+}+P_{\varepsilon_{ij}}^{(4)+}$ \eqref{Eq_DTWT_s_epsijBs_ss_epsijTEq_002},
which behave differently for each component across the channel \figref{Fig_DTWT_s_epsijBs_ss_epsijTBs_002}.
Globally, in plane channel flow, $P_{\varepsilon_{ij}}^+$ contributes as a gain mechanism in the $\varepsilon_{ij}$-budgets \figrefsab{Fig_DTWT_s_epsijBs_ss_epsijTBs_001}
                                                                                                                                      {Fig_DTWT_s_epsijBs_ss_epsijTBs_002},
except very near the wall for the wall-normal ($P_{\varepsilon_{yy}}^+\leq0\;\forall\;y^+\lessapprox3$ is a loss mechanism; \figrefnp{Fig_DTWT_s_epsijBs_ss_epsijTBs_001})
and spanwise ($P_{\varepsilon_{zz}}^+\leq0\;\forall\;y^+\lessapprox2$ is a loss mechanism; \figrefnp{Fig_DTWT_s_epsijBs_ss_epsijTBs_001}) diagonal components.
For both the spanwise and wall-normal components, $P_{\varepsilon_{yy}}^{(1)+}=P_{\varepsilon_{zz}}^{(1)+}=P_{\varepsilon_{yy}}^{(3)+}=P_{\varepsilon_{zz}}^{(3)+}=0\;\forall\;y^+$ \tabref{Tab_DTWT_s_epsijBs_ss_epsijTBs_001},
whereas production by the action of $\mathcal{E}_{ijkm}^+$-components on the mean velocity-gradient, $P_{\varepsilon_{yy}}^{(2)+}$ and 
$P_{\varepsilon_{zz}}^{(2)+}$ \tabref{Tab_DTWT_s_epsijBs_ss_epsijTBs_001}, is the main gain mechanism in the initial part of the buffer layer ($y^+\in [5,30]$; \figrefnp{Fig_DTWT_s_epsijBs_ss_epsijTBs_002})
and is replaced further away from the wall ($y^+\gtrapprox 30$; \figrefnp{Fig_DTWT_s_epsijBs_ss_epsijTBs_002}) by production by triple correlations of fluctuating velocity-gradients, $P_{\varepsilon_{yy}}^{(4)+}$ and $P_{\varepsilon_{zz}}^{(4)+}$
\eqref{Eq_DTWT_s_epsijBs_ss_epsijTEq_002}, as expected from the quasi-homogeneous analysis of \citet[pp. 88--92]{Tennekes_Lumley_1972a}.
All of the 4 production mechanisms \eqref{Eq_DTWT_s_epsijBs_ss_epsijTEq_002} are active for the streamwise $\varepsilon_{xx}^+$ and shear $\varepsilon_{xy}^+$ components \figref{Fig_DTWT_s_epsijBs_ss_epsijTBs_002}.
Regarding the streamwise component $\varepsilon_{xx}^+$, the 3 mechanisms, $P_{\varepsilon_{xx}}^{(1)+}>0$, $P_{\varepsilon_{xx}}^{(2)+}>0$ and $P_{\varepsilon_{xx}}^{(4)+}>0$ $\forall\;y^+>0$ \figref{Fig_DTWT_s_epsijBs_ss_epsijTBs_002}
always contribute as gain to the budgets.
On the other hand, production by the mean velocity-Hessian, $P_{\varepsilon_{xx}}^{(3)+}<0 \;\forall\;y^+\lessapprox 10$ (loss) near the wall \figref{Fig_DTWT_s_epsijBs_ss_epsijTBs_002}, switching to $P_{\varepsilon_{xx}}^{(3)+}>0 \;\forall\;y^+\gtrapprox 10$ (gain)
further away from the wall, is generally weaker than the other 3 mechanisms. Notice that the production by the triple correlations of the fluctuating velocity-gradients $P_{\varepsilon_{xx}}^{(4)+}$ \eqref{Eq_DTWT_s_epsijBs_ss_epsijTEq_002}
is important throughout the channel, even as $y^+\to 0$ \figref{Fig_DTWT_s_epsijBs_ss_epsijTBs_002}, becoming the dominant mechanism in the outer part of the flow ($y^+\gtrapprox 15$; \figrefnp{Fig_DTWT_s_epsijBs_ss_epsijTBs_002}).
Regarding the shear component $\varepsilon_{xy}^+$, $P_{\varepsilon_{xy}}^{(1)+}=-\varepsilon_{yy}^+[d_y\bar  u]^+<0\;\forall\;y^+>0$ \figref{Fig_DTWT_s_epsijBs_ss_epsijTBs_002}
is the main gain mechanism throughout the channel, whereas $P_{\varepsilon_{xy}}^{(2)+}>0$ and $P_{\varepsilon_{xy}}^{(3)+}>0\;\forall\;y^+>0$ contribute as loss to the budgets \figref{Fig_DTWT_s_epsijBs_ss_epsijTBs_002}.
Finally, $P_{\varepsilon_{xy}}^{(4)+}<0\,\forall y^+\lessapprox6$ (gain) near the wall \figref{Fig_DTWT_s_epsijBs_ss_epsijTBs_002} switches to $P_{\varepsilon_{xy}}^{(4)+}>0\;\forall\;y^+\gtrapprox6$ (loss) further away from the wall.
Interestingly, $P_{\varepsilon_{xy}}^{(4)+}$, although comparable to the other 3 mechanisms $\forall\;y^+$ \figref{Fig_DTWT_s_epsijBs_ss_epsijTBs_002}, never becomes the dominant mechanism, even as $y^+\to \delta^+$, raising
the question of applicability of the quasi-homogeneous order-of-magnitude analysis \citep[pp. 88--92]{Tennekes_Lumley_1972a} to the shear component (although higher-$Re$ data are required to fully resolve this issue).

In plane channel flow, production by interaction with the mean velocity-gradient $P_{\varepsilon_{ij}}^{(1)+}+P_{\varepsilon_{ij}}^{(2)+}$ \eqref{Eq_DTWT_s_epsijBs_ss_epsijTEq_002}
involves $\varepsilon_{xy}^+$ in $P_{\varepsilon_{xx}}^{(1)+}$ \tabref{Tab_DTWT_s_epsijBs_ss_epsijTBs_001},
$\varepsilon_{yy}^+$ in $P_{\varepsilon_{xy}}^{(1)+}$ \tabref{Tab_DTWT_s_epsijBs_ss_epsijTBs_001}, and 5 different components of $\mathcal{E}_{ijkm}$ \eqref{Eq_DTWT_s_epsijBs_001} in $P_{\varepsilon_{ij}}^{(2)+}$ \tabref{Tab_DTWT_s_epsijBs_ss_epsijTBs_001},
which are not involved in the generating relation \eqref{Eq_DTWT_s_epsijBs_001a}.
Since the mean velocity-gradient $d_y\bar  u^+>0\,\forall y^+\in]0,\delta^+[$ \citep{Coles_1956a}, the sign of these components directly determines \tabref{Tab_DTWT_s_epsijBs_ss_epsijTBs_001} whether the corresponding contribution to $P_{\varepsilon_{ij}}^+$
is gain or loss \figrefsab{Fig_DTWT_s_epsijBs_ss_epsijTBs_002}
                          {Fig_DTWT_s_epsijBs_ss_epsijTBs_003}.

Concerning $P_{\varepsilon_{xx}}^{(1)+}$ and $P_{\varepsilon_{xx}}^{(2)+}$ \tabref{Tab_DTWT_s_epsijBs_ss_epsijTBs_001}, the corresponding producing components, $\varepsilon_{xy}^+$ and $\mathcal{E}_{xxxy}^+$ are of comparable
magnitude \figref{Fig_DTWT_s_epsijBs_ss_epsijTBs_003}, but $\varepsilon_{xy}^+$ is active nearer to the wall (peak at $y^+\in[4,5]$; \figrefnp{Fig_DTWT_s_epsijBs_ss_epsijTBs_003})
compared to $\mathcal{E}_{xxxy}^+$ (peak at $y^+\approxeq 15$; \figrefnp{Fig_DTWT_s_epsijBs_ss_epsijTBs_003}). Therefore, $P_{\varepsilon_{xx}}^{(1)+}$ peaks at $y^+\in[4,5]$ whereas $P_{\varepsilon_{xx}}^{(2)+}$ peaks at $y^+\approxeq 9$
\figref{Fig_DTWT_s_epsijBs_ss_epsijTBs_002}. Near the wall (wall-asymptotic expansion; \parrefnp{DTWT_s_DNSDepsij_ss_WA}), $P_{\varepsilon_{xx}}^{(2)+}{\underset{y^+\to0}{\sim}} O(y^{+2})$
whereas $P_{\varepsilon_{xx}}^{(1)+}{\underset{y^+\to0}{\sim}} O(y^{+})$ \figref{Fig_DTWT_s_epsijBs_ss_epsijTBs_002}. For the wall-normal and spanwise diagonal components $P_{\varepsilon_{yy}}^{(1)+}=P_{\varepsilon_{zz}}^{(1)+}=0$
\tabref{Tab_DTWT_s_epsijBs_ss_epsijTBs_001}, so that only $P_{\varepsilon_{yy}}^{(2)+}$ and $P_{\varepsilon_{zz}}^{(2)+}$ appear in the budgets of $\varepsilon_{yy}^+$ and $\varepsilon_{zz}^+$
respectively \figref{Fig_DTWT_s_epsijBs_ss_epsijTBs_002}, with corresponding producing components $\mathcal{E}_{yyxy}^+$ and $\mathcal{E}_{zzxy}^+$ respectively \tabref{Tab_DTWT_s_epsijBs_ss_epsijTBs_001}.
These components peak at $y^+\approxeq 30$ ($\mathcal{E}_{yyxy}^+$; \figrefnp{Fig_DTWT_s_epsijBs_ss_epsijTBs_003}) and $y^+\approxeq 25$ ($\mathcal{E}_{zzxy}^+$; \figrefnp{Fig_DTWT_s_epsijBs_ss_epsijTBs_003}),
the $\mathcal{E}_{zzxy}^+$-peak being larger than the $\mathcal{E}_{yyxy}^+$-peak \figref{Fig_DTWT_s_epsijBs_ss_epsijTBs_003}. The producing component $\mathcal{E}_{yyxy}^+\lessapprox0\,\forall\,y^+>0$ \figref{Fig_DTWT_s_epsijBs_ss_epsijTBs_003}, so that
\tabref{Tab_DTWT_s_epsijBs_ss_epsijTBs_001} $P_{\varepsilon_{yy}}^{(2)+}\gtrapprox0\,\forall\,y^+>0$ (gain; \figrefnp{Fig_DTWT_s_epsijBs_ss_epsijTBs_002}) with a peak at $y^+\approxeq 18$.
On the other hand, $\mathcal{E}_{zzxy}^+<0\,\forall\,y^+\gtrapprox 2$ (gain; \figrefnp{Fig_DTWT_s_epsijBs_ss_epsijTBs_003}) changes sign near the wall ($\mathcal{E}_{zzxy}^+>0\,\forall\,y^+\lessapprox 2$; loss; \figrefnp{Fig_DTWT_s_epsijBs_ss_epsijTBs_003}).
Regarding the shear component $\varepsilon_{xy}^+$, the producing terms $\varepsilon_{yy}^+>0\,\forall\,y^+>0$ \figref{Fig_DTWT_s_epsijBs_ss_epsijTBs_003} in $P_{\varepsilon_{xy}}^{(1)+}$ \tabref{Tab_DTWT_s_epsijBs_ss_epsijTBs_001}
and $(\mathcal{E}_{xyxy}^+ + \mathcal{E}_{xyyx}^+)\lessapprox 0 \,\forall \, y^+>0$ \figref{Fig_DTWT_s_epsijBs_ss_epsijTBs_003} in $P_{\varepsilon_{xy}}^{(2)+}$ \tabref{Tab_DTWT_s_epsijBs_ss_epsijTBs_001}, so that
$P_{\varepsilon_{xy}}^{(2)+}\gtrapprox 0\,\forall \, y^+>0$ (loss; \figrefnp{Fig_DTWT_s_epsijBs_ss_epsijTBs_002}) opposes, but is weaker than, $P_{\varepsilon_{xy}}^{(1)+}\lessapprox 0\,\forall \, y^+>0$ (gain; \figrefnp{Fig_DTWT_s_epsijBs_ss_epsijTBs_002}).
\begin{figure}
\begin{center}
\begin{picture}(450,535)
\put(0,0){\includegraphics[angle=0,width=380pt,bb=59 101 516 742]{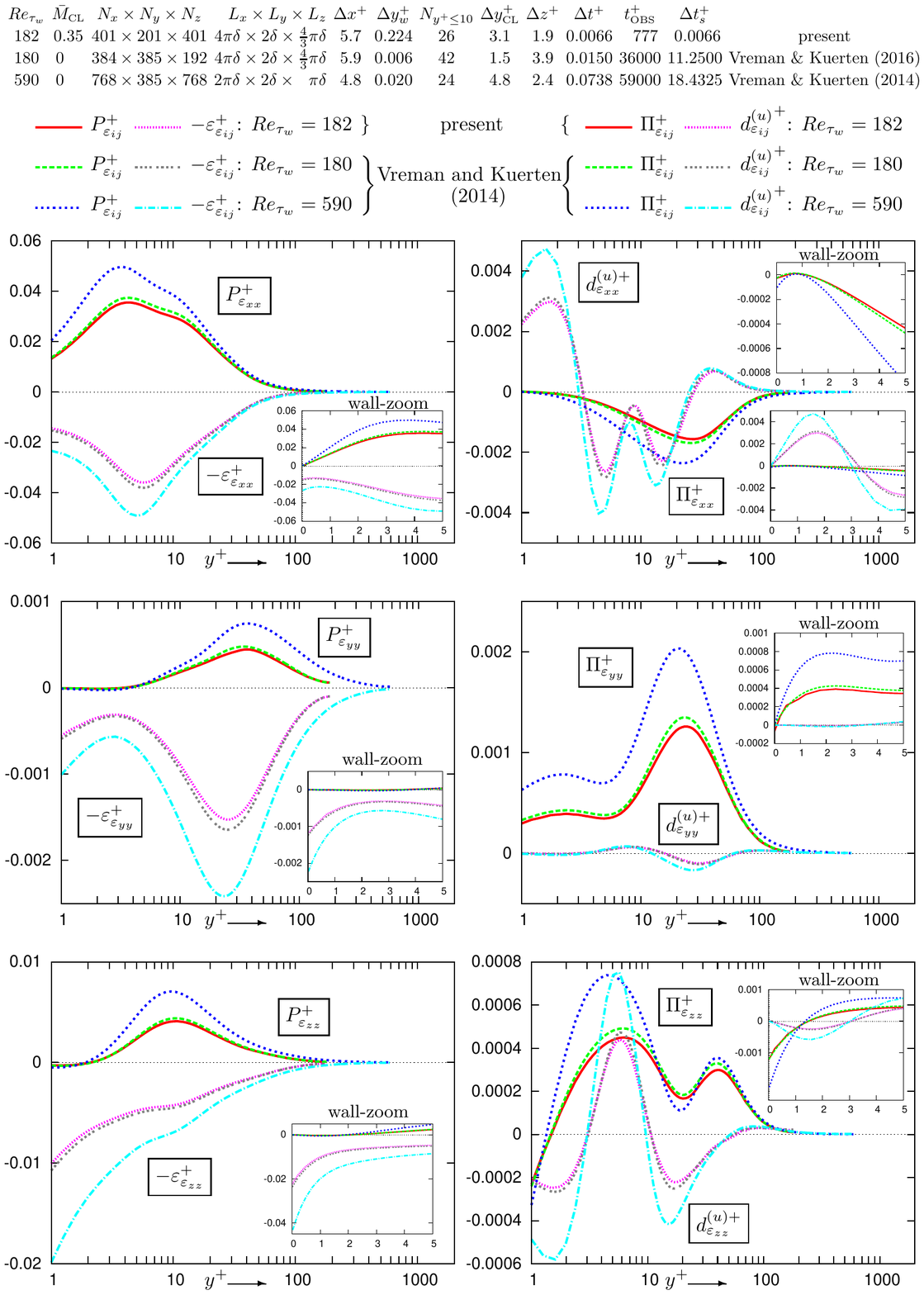}}
\end{picture}
\end{center}
\caption{Examination of the influence of the Reynolds number $Re_{\tau_w}$ \eqref{Eq_DTWT_s_AppendixABVSy+0_ss_WUs_001g}
on various terms ($P_{\varepsilon_{ij}}$, $\varepsilon_{\varepsilon_{ij}}$, $\Pi_{\varepsilon_{ij}}$, and $d^{(u)}_{\varepsilon_{ij}}$)
in the budgets of the transport equations \eqref{Eq_DTWT_s_epsijBs_ss_epsijTEq_002}
for the diagonal components of the dissipation tensor $\varepsilon_{ij}$ \eqref{Eq_DTWT_s_I_002a}, in wall-units \eqref{Eq_DTWT_s_AppendixABVSy+0_ss_WUs_001f},
by comparison of the present \tsn{DNS} computations of turbulent plane channel flow ($Re_{\tau_w}\approxeq182$, $\bar M_\tsn{CL}\approxeq0.35$) with the incompressible \tsn{DNS} data
of \citet[$Re_{\tau_w}\in\{180,590\}$, $\bar M_\tsn{CL}=0$]{Vreman_Kuerten_2014a,
                                                            Vreman_Kuerten_2014b,
                                                            Vreman_Kuerten_2016a}
plotted against the inner-scaled \eqref{Eq_DTWT_s_AppendixABVSy+0_ss_WUs_001c} wall-distance $y^+$ (logscale and linear wall-zoom).}
\label{Fig_DTWT_s_RetauwIoepsijBs_001}
\end{figure}
%
\begin{figure}
\begin{center}
\begin{picture}(450,535)
\put(0,0){\includegraphics[angle=0,width=380pt,bb=60 101 520 742]{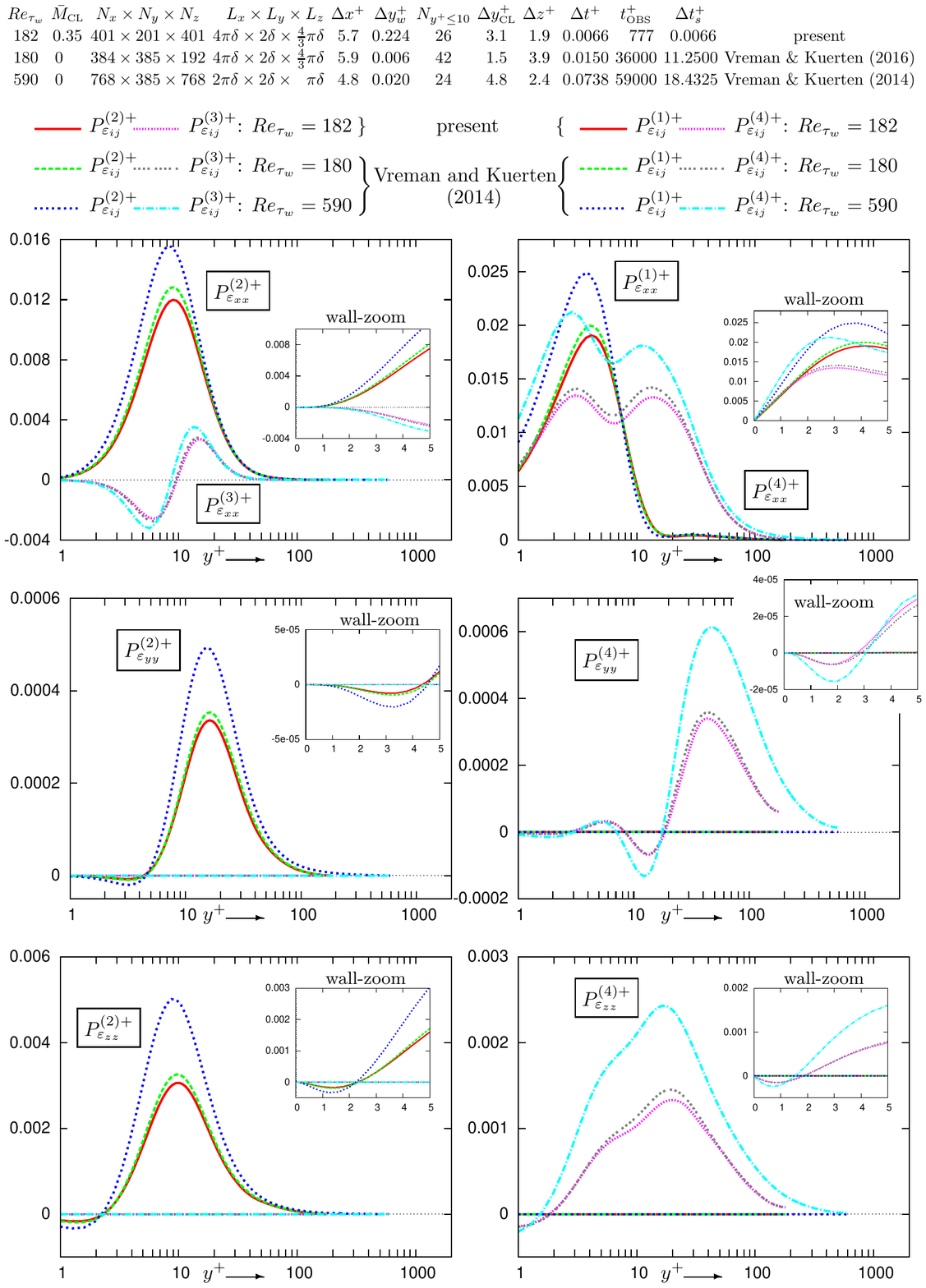}}
\end{picture}
\end{center}
\caption{Examination of the influence of the Reynolds number $Re_{\tau_w}$ \eqref{Eq_DTWT_s_AppendixABVSy+0_ss_WUs_001g}
on the of production $P_{\varepsilon_{ij}}=P^{(1)}_{\varepsilon_{ij}}+P^{(2)}_{\varepsilon_{ij}}+P^{(3)}_{\varepsilon_{ij}}+P^{(4)}_{\varepsilon_{ij}}$ \eqref{Eq_DTWT_s_epsijBs_ss_epsijTEq_002}
of the diagonal components of the dissipation tensor $\varepsilon_{ij}$ \eqref{Eq_DTWT_s_I_002a}, in wall-units \eqref{Eq_DTWT_s_AppendixABVSy+0_ss_WUs_001f},
by comparison of the present \tsn{DNS} computations of turbulent plane channel flow ($Re_{\tau_w}\approxeq182$, $\bar M_\tsn{CL}\approxeq0.35$) with the incompressible \tsn{DNS} data
of \citet[$Re_{\tau_w}\in\{180,590\}$, $\bar M_\tsn{CL}=0$]{Vreman_Kuerten_2014a,
                                                            Vreman_Kuerten_2014b,
                                                            Vreman_Kuerten_2016a}
plotted against the inner-scaled \eqref{Eq_DTWT_s_AppendixABVSy+0_ss_WUs_001c} wall-distance $y^+$ (logscale and linear wall-zoom).}
\label{Fig_DTWT_s_RetauwIoepsijBs_002}
\end{figure}
%
%
%
%
%
%
%
%
%
\section{$Re_{\tau_w}$ influence on $\varepsilon_{ij}$-budgets}\label{DTWT_s_RetauwIoepsijBs}
%
%
%
%
%
%
%
%
%

The previous analysis of $\varepsilon_{ij}$ \parref{DTWT_s_epsijBs} concerned the low $Re_{\tau_w}\approxeq180$ case. 
It turns out that the data of \citet{Vreman_Kuerten_2014a,Vreman_Kuerten_2014b,Vreman_Kuerten_2016a} for the budgets of the transport equations for the 9 components of the velocity-gradient variance $\overline{(\partial_{x_j}u_i')^2}$ 
\citep[(9--18), p.~4]{Vreman_Kuerten_2014b} can be combined to yield the budgets of the transport equations of the diagonal components $\{\varepsilon_{xx}^+,\varepsilon_{yy}^+,\varepsilon_{zz}^+\}$ for $Re_{\tau_w}\in \{180,590\}$
\figrefsab{Fig_DTWT_s_RetauwIoepsijBs_001}{Fig_DTWT_s_RetauwIoepsijBs_002}. Unfortunately, data for the budgets of the transport equation for the shear component $\varepsilon_{xy}^+$ could not be obtained from this database.
The data of \citet{Vreman_Kuerten_2014a,Vreman_Kuerten_2014b,Vreman_Kuerten_2016a} were sampled at much lower frequencies ($\Delta t_s^+\in\{11.25,18.43\}$) but for much longer observation 
times ($t_{\tsn{OBS}}^+\in\{36000,59000\}$).

With regard to the lower Reynolds number $Re_{\tau_w}\approxeq180$, the agreement between the present $\bar M_\tsn{CL}\approxeq0.35$ aerodynamic \tsn{DNS} data and the incompressible \tsn{DNS} data
of \citet{Vreman_Kuerten_2016a} is excellent, the 2 sets of data practically collapsing on the same curves \figrefsab{Fig_DTWT_s_RetauwIoepsijBs_001}
                                                                                                                                       {Fig_DTWT_s_RetauwIoepsijBs_002}.
This agreement between completely different computational approaches and flow models \citep{Gerolymos_Senechal_Vallet_2010a,
                                                                                            Vreman_Kuerten_2016a},
spatial (in particular $y$-wise) and temporal resolutions, and averaging times \figrefsab{Fig_DTWT_s_RetauwIoepsijBs_001}
                                                                                         {Fig_DTWT_s_RetauwIoepsijBs_002},
both strengthens considerably the confidence in the data and corroborates the analysis of weak-density-fluctuation effects \citep[Appendix A, pp. 45--51]{Gerolymos_Senechal_Vallet_2013a},
showing that compressibility effects on turbulence structure are indeed negligible for $\bar M_\tsn{CL}\lessapprox0.35$, in line with the finding that $\rho'_\mathrm{rms}\propto\bar\rho\bar M_\tsn{CL}^2$
\citep[Fig. 5, p. 720]{Gerolymos_Vallet_2014a}.

Regarding the influence of $Re_{\tau_w}$ on the diagonal components of various terms ($P_{\varepsilon_{ij}}$, $\varepsilon_{\varepsilon_{ij}}$, $\Pi_{\varepsilon_{ij}}$, and $d^{(u)}_{\varepsilon_{ij}}$)
in the $\varepsilon_{ij}$-budgets \eqref{Eq_DTWT_s_epsijBs_ss_epsijTEq_002},
although the level of wall-values and of different peaks present in the wall-normal ($y$-wise) distributions are higher with increasing Reynolds number \figrefsab{Fig_DTWT_s_RetauwIoepsijBs_001}
                                                                                                                                                                  {Fig_DTWT_s_RetauwIoepsijBs_002},
there are no significant qualitative differences. However, the locations and/or values of some near-wall extrema exhibit substantial variation as $Re_{\tau_w}$ increases from $180$ to $590$.
These variations are analogous to the observed behaviour of near-wall anisotropy \figrefsatob{Fig_DTWT_s_DNSDepsij_ss_RetauwIA_sss_ATs_001}
                                                                                             {Fig_DTWT_s_DNSDepsij_ss_RetauwIA_sss_AIs_001},
but this analogy also suggests that lesser influence should be expected as $Re_{\tau_w}$ further increases above $590$.
Regarding the location of near-wall extrema of different terms there is a general trend \figrefsab{Fig_DTWT_s_RetauwIoepsijBs_001}
                                                                                                  {Fig_DTWT_s_RetauwIoepsijBs_002}
that they occur at lower $y^+$ (closer to the wall) as $Re_{\tau_w}$ increases from $180$ to $590$. Notice, however, that the very-near-wall location where the streamwise and spanwise components of
the destruction-of-dissipation tensor $\varepsilon_{\varepsilon_{xx}}=\varepsilon_{\varepsilon_{zz}}$, exhibits the opposite behaviour, very slightly increasing from $y^+\approxeq0.6$ at $Re_{\tau_w}=180$
to $y^+\approxeq0.8$ at $Re_{\tau_w}=590$. Another noticeable $Re_{\tau_w}$-effect is the difference of the near-wall levels of the wall-normal and spanwise components
of the destruction-of-dissipation tensor, $\varepsilon_{\varepsilon_{yy}}$ and $\varepsilon_{\varepsilon_{zz}}$, which are at $Re_{\tau_w}=590$ nearly twice those observed at $Re_{\tau_w}=180$
\figref{Fig_DTWT_s_RetauwIoepsijBs_001}.

%
%
%
%
%
%
%
%
%
\section{Conclusions}\label{DTWT_s_C}
%
%
%
%
%
%
%
%
%

Available and novel \tsn{DNS} data were used to study the positive-definite dissipation tensor $\varepsilon_{ij}$ \eqref{Eq_DTWT_s_I_002a},
representing the destruction of the Reynolds-stresses $r_{ij}$ \eqref{Eq_DTWT_s_DNSDepsij_ss_A_001a} by the action of molecular viscosity, in wall turbulence,
and in particular its anisotropy and transport-equations budgets.

Taylor-expansions of the fluctuating velocities in the neighbourhood of a plane no-slip $xz$-wall show that, for incompressible turbulent flow,
the wall-normal gradients of the $\varepsilon_{ij}$-anisotropy tensor $b_{\varepsilon_{ij}}$ and of its
invariants, at the wall, are exactly twice the wall-normal gradients of the corresponding conponents and invariants of the Reynolds-stress anisotropy tensor $b_{ij}$. Furtheremore,
both the dissipation tensor $\varepsilon_{ij}$ and the
Reynolds-stress tensor $r_{ij}$, depart from the 2-C state at the wall (flatness parameter $A_w=A_{\varepsilon_w}=0$) quadratically with wall-distance $y^+$
(\smash{$A_\varepsilon\sim_{y^+\to0}4A\sim_{y^+\to0}O({y^+}^2)$}).

Available \tsn{DNS} data suggest several general trends in the anisotropy of $r_{ij}$ and $\varepsilon_{ij}$, with varying Reynolds number $Re_{\tau_w}$.
The $y^+$-wise distributions of the components and invariants of the Reynolds-stress anisotropy tensor $b_{ij}$ develop, as $Re_{\tau_w}$ increases, a plateau, roughly corresponding to the log-layer of the mean velocity profile,
followed by a wake-like region in the outer part near the centerline. At the centerline, \tsn{DNS} data show quite consistently that $b_{ij}$ reaches a rod-like axisymmetric componentality, at least for $Re_{\tau_w}\gtrapprox180$.
On the contrary, the components and invariants of $b_{\varepsilon_{ij}}$ seem, as $Re_{\tau_w}$ increases, to vary smoothly from $y^+\approxeq 100$ to the centerline, where $\varepsilon_{ij}$ is not axisymmetric. Interestingly, at the low
$Re_{\tau_w}$ limit, the anisotropy invariants ($-\II{b}$, $\III{b}$, $-\II{b_\varepsilon}$, $\III{b_\varepsilon}$) seem to increase continuously, with decreasing $Re_{\tau_w}$, both at the wall and at the centerline.

The dissipation tensor $\varepsilon_{ij}$ \eqref{Eq_DTWT_s_I_002a}, is governed by transport equations \eqref{Eq_DTWT_s_epsijBs_ss_epsijTEq_002} where convection by the mean-flow $C_{\varepsilon_{\ij}}$ is balanced by the usual mechanisms:
molecular diffusion $d_{\varepsilon_{ij}}^{(\mu)}$,
turbulent diffusion $d_{\varepsilon_{ij}}^{(u)}$,
production $P_{\varepsilon_{ij}}$
the effect of the fluctuating pressure-Hessian $\Pi_{\varepsilon_{ij}}$, and
destruction by molecular viscosity $\varepsilon_{\varepsilon_{ij}}$.

Budgets for these equations were studied using \tsn{DNS} results for low $Re_{\tau_w}\approxeq180$ turbulent plane channel flow (for this particular flow convection $C_{\varepsilon_{\ij}}=0$).
As expected, since $\varepsilon_{ij}$ is the footprint of the behaviour of the small turbulent scales, the various mechanisms in the $\varepsilon_{ij}$-budgets behave unlike
the corresponding mechanisms in the $r_{ij}$-budgets \citep{Mansour_Kim_Moin_1988a}.
Production $P_{\varepsilon_{ij}}$ is significant (gain) for all of the $\varepsilon_{ij}$-components, contrary to the $r_{ij}$-budgets, where for plane channel flow
$P_{yy}=P_{zz}=0\;\forall\;y^+$. On the other hand, the pressure term $\Pi_{\varepsilon_{ij}}$ is significant in the budgets of the wall-normal ($\Pi_{\varepsilon_{yy}}$)
and shear ($\Pi_{\varepsilon_{xy}}$) components and negligibly small in the budgets of the streamwise ($\Pi_{\varepsilon_{xx}}$) and spanwise ($\Pi_{\varepsilon_{zz}}$) components.
This contrasts with the $r_{ij}$-budgets, where $\Pi_{zz}$ is the main gain mechanism in the ($r_{zz}:=\overline{w'^2}$)-budgets
and $\Pi_{xx}$, although relatively weak near the wall, is an important loss mechanism for $y^+\gtrapprox20$ in the ($r_{xx}:=\overline{u'^2}$)-budgets \citep{Mansour_Kim_Moin_1988a}.
Production of the diagonal components ($\varepsilon_{xx}$, $\varepsilon_{yy}$ and $\varepsilon_{zz}$) by the triple correlations of the fluctuating velocity-gradients $P_{\varepsilon_{ij}}^{(4)}$
is the main gain mechanism, in line with quasi-homogeneous theory \citep[pp. 88--92]{Tennekes_Lumley_1972a}, only away from the wall ($y^+\gtrapprox20$),
but this does not apply for the shear component $\varepsilon_{xy}$, for which $P_{\varepsilon_{xy}}^{(4)}$ is a loss mechanism in the major part of the channel ($y^+\gtrapprox6$).
The budgets of the diagonal components ($\varepsilon_{xx}$, $\varepsilon_{yy}$ and $\varepsilon_{zz}$) were also evaluated at higher $Re_{\tau_w}\approxeq590$,
by combining available \tsn{DNS} data of \citet{Vreman_Kuerten_2014b} for the variances of the fluctuating velocity-gradient components,
showing both the same qualitative behaviour as for the lower $Re_{\tau_w}\approxeq180$ case and higher values of the different peaks.

In addition to the ubiquitous effort to extend the \tsn{DNS} results to higher Reynolds numbers, there are several new research directions suggested by the present results,
which are the subject of ongoing work: (a) investigate the dynamics and transport-equation budgets of the destruction-of-dissipation tensor
$\varepsilon_{\varepsilon_{ij}}$ in an effort to understand the very-small-scale inhomogeneity information it represents, (b) complete the very-low $Re_{\tau_w}\lessapprox100$ range of \tsn{DNS} data to further substantiate and investigate the specific
anisotropy behaviour observed in the \citet{Hu_Morfey_Sandham_2002a,
                                           Hu_Morfey_Sandham_2003a,
                                           Hu_Morfey_Sandham_2006a} channel data, and (c) to exploit the \tsn{DNS} database for the development and assessment of complete $r_{ij}$--$\varepsilon_{ij}$ second-moment closures,
which by including transport equations for all the components of the lengthscale tensor are better adapted to the strong inhomogeneity-induced anisotropy of practical turbulent wall-bounded flows.

\begin{acknowledgments}
The authors are listed alphabetically.
The present work was partly funded by the \tsn{ANR} project \NumERICCS (\tsn{ANR--15--CE06--0009}).
Computations were performed using \tsn{HPC} ressources allocated at \tsn{GENCI--IDRIS} (Grant 2015--022139)
and at \tsn{ICS--UPMC} (\tsn{ANR--10--EQPX--29--01}).
Tabulated data are available at {\tt http://www.aerodynamics.fr/DNS\_database/CT\_chnnl}.
\end{acknowledgments}

%
%
%
%
%
%
%
%
%
\oneappendix\section{Asymptotic behaviour in the viscous sublayer ($y^+\to0$)}\label{DTWT_s_AppendixABVSy+0}
%
%
%
%
%
%
%
%
%

Assume turbulent flow near a plane wall, coincident with the $xz$-plane and located at $y^+=0$.
Near the wall, all fluctuating quantities are expandend $y$-wise in Taylor-series around $y^+=0$
\begin{alignat}{6}
(\cdot)'^+\underset{y^+\to0}{\sim}(\cdot)_w'^+(x^+,z^+,t^+)&+&A_{(\cdot)}'^+(x^+,z^+,t^+)\;y^+    &+&B_{(\cdot)}'^+(x^+,z^+,t^+)\;{y^+}^2& &
                                                                                                                                    \notag\\
                                                           &+&C_{(\cdot)}'^+(x^+,z^+,t^+)\;{y^+}^3&+&D_{(\cdot)}'^+(x^+,z^+,t^+)\;{y^+}^4&+&\cdots
                                                                                                                                    \label{Eq_DTWT_s_AppendixABVSy+0_001}
\end{alignat}
with coefficients proportional to the wall-normal ($y$) derivatives of the appropriate order, which are stationary random functions of $\{x^+,z^+,t^+\}$.
The limiting behaviour in the viscous sublayer is determined by the no-slip condition at the wall \cite{Mansour_Kim_Moin_1988a}.
\begin{alignat}{6}
&y^+\in\{0,2\delta^+\}\;\implies\;\bar u^+=\bar v^+=\bar w^+=u'^+=v'^+=w'^+=0\quad;\quad\forall x^+,z^+,t^+
                                                                                                                                    \label{Eq_DTWT_s_AppendixABVSy+0_002}
\end{alignat}

%
%
%
%
%
\subsection{Wall units}\label{DTWT_s_AppendixABVSy+0_ss_WUs}
%
%
%
%
%

All variables are made nondimensional using the mean wall shear stress, the constant fluid density $\rho$ and the constant dynamic viscosity $\nu$
\begin{subequations}
                                                                                                                                    \label{Eq_DTWT_s_AppendixABVSy+0_ss_WUs_001}
\begin{alignat}{6}
\bar\tau_w:=[\bar\tau_{xy}]_w\qquad;\qquad
\rho\approxeq\const\qquad;\qquad
\nu\approxeq\const
                                                                                                                                    \label{Eq_DTWT_s_AppendixABVSy+0_ss_WUs_001a}
\end{alignat}
which define the friction-velocity
\begin{alignat}{6}
u_\tau:=\sqrt{\dfrac{\bar\tau_w}{\rho}}
                                                                                                                                    \label{Eq_DTWT_s_AppendixABVSy+0_ss_WUs_001b}
\end{alignat}
where $\tau_{ij}$ is the viscous stress tensor \citep[(2.4), p. 31]{Davidson_2004a}.
Using wall-units \eqrefsab{Eq_DTWT_s_AppendixABVSy+0_ss_WUs_001a}
                          {Eq_DTWT_s_AppendixABVSy+0_ss_WUs_001b},
the nondimensional variables $(\cdot)^+$ are defined as
\begin{alignat}{6}
y^+:=  \dfrac{u_\tau\;(y-y_w)}{\nu}             \quad;\quad
t^+:=& \dfrac{t\;u_\tau^2}{\nu}                 \quad;\quad
u_i^+:=\dfrac{u_i}{u_\tau}
                                                                                                                                    \label{Eq_DTWT_s_AppendixABVSy+0_ss_WUs_001c}\\
\left[r_{ij}^+,\tau_{ij}^+,p^+\right]^\tsn{T}:=&\dfrac{1}{\rho u_\tau^2}\left[\rho\overline{u_i'u_j'},\tau_{ij},p\right]^\tsn{T}
                                                                                                                                    \label{Eq_DTWT_s_AppendixABVSy+0_ss_WUs_001d}\\
\left[\varepsilon^+_{ij}, P_{ij}^+,\Pi_{ij}^+,d_{ij}^+\right]^\tsn{T}:=&\dfrac{\nu}{\rho\;u_\tau^4}\left[\rho\varepsilon_{ij},P_{ij},\Pi_{ij},d_{ij}\right]^\tsn{T}
                                                                                                                                    \label{Eq_DTWT_s_AppendixABVSy+0_ss_WUs_001e}\\
\left[\varepsilon^+_{\varepsilon_{ij}},P_{\varepsilon_{ij}}^+,\Pi_{\varepsilon_{ij}}^+,d_{\varepsilon_{ij}}^+\right]^\tsn{T}:=&
\dfrac{\nu^2}{\rho\;u_\tau^6}\left[\rho\varepsilon_{\varepsilon_{ij}},P_{\varepsilon_{ij}},\Pi_{\varepsilon_{ij}},d_{\varepsilon_{ij}}\right]^\tsn{T}
                                                                                                                                    \label{Eq_DTWT_s_AppendixABVSy+0_ss_WUs_001f}
\end{alignat}
\ie\ terms in $r_{ij}$-transport \eqref{Eq_DTWT_s_I_001} scale as $\rho\nu^{-1}u_\tau^4$
whereas terms in $\varepsilon_{ij}$-transport \eqref{Eq_DTWT_s_epsijBs_ss_epsijTEq_002} scale as $\rho\nu^{-2}u_\tau^6$ \citep[p. 42]{Jovanovic_2004a}.
In fully developed turbulent channel flow, the friction Reynolds number is defined as
\begin{alignat}{6}
Re_{\tau_w}:=\dfrac{u_\tau\;\delta}{\nu}\stackrel{\eqref{Eq_DTWT_s_AppendixABVSy+0_ss_WUs_001c}}{=}\delta^+
                                                                                                                                    \label{Eq_DTWT_s_AppendixABVSy+0_ss_WUs_001g}
\end{alignat}
where $2\delta$ is the channel's height.
\end{subequations}

%
%
%
%
%
\subsection{Anisotropy tensors and invariants}\label{DTWT_s_AppendixABVSy+0_ss_ATsIs}
%
%
%
%
%

The expansions \eqref{Eq_DTWT_s_DNSDepsij_ss_WA_001}
can be used to obtain the expansions for the anisotropy tensors $b_{ij}$ \eqref{Eq_DTWT_s_DNSDepsij_ss_A_001a} and
$b_{\varepsilon_{ij}}$ \eqref{Eq_DTWT_s_DNSDepsij_ss_A_002a},
and their invariants \eqrefsab{Eq_DTWT_s_DNSDepsij_ss_A_001b}
                              {Eq_DTWT_s_DNSDepsij_ss_A_002b}.
Recall that if 
\begin{subequations}
                                                                                                                                    \label{Eq_DTWT_s_AppendixABVSy+0_ss_ATsIs_001}
\begin{alignat}{6}
P(x)\sim\sum_{m=0}^{\infty}\alpha_m\;x^m\qquad;\qquad
Q(x)\sim\sum_{m=0}^{\infty}\beta_m\;x^m
                                                                                                                                    \label{Eq_DTWT_s_AppendixABVSy+0_ss_ATsIs_001a}
\end{alignat}
then the asymptotic expansion of their ratio is given by
\begin{alignat}{6}
\dfrac{P(x)}{Q(x)}\sim\sum_{\ell=0}^{\infty}\gamma_\ell\;x^\ell\iff&
\left(\sum_{\ell=0}^{\infty}\gamma_\ell\;x^\ell\right)\left(\sum_{m=0}^{\infty}\beta_m\;x^m\right)\sim\sum_{n=0}^{\infty}\alpha_n\;x^n
                                                                                                                                    \notag\\
\iff&
\sum_{n=0}^{\infty}\left(\left(\sum_{\ell=0}^{n}\beta_{n-\ell}\gamma_\ell\right)-\alpha_n\right)\;x^n\sim0
                                                                                                                                    \label{Eq_DTWT_s_AppendixABVSy+0_ss_ATsIs_001b}
\end{alignat}
\end{subequations}
resulting in a series of linear relations that can be solved sequentially to obtain the coefficients $\gamma_n$.
Straightforward calculations yield the wall-asymptotic expansions of the anisotropy tensors
$\tsr{b}$ \tabref{Tab_DTWT_s_DNSDepsij_ss_WA_001} and $\tsr{b_\varepsilon}$ \tabref{Tab_DTWT_s_DNSDepsij_ss_WA_002}.
Substitution of these expansions in \eqrefsab{Eq_DTWT_s_DNSDepsij_ss_A_001b}{Eq_DTWT_s_DNSDepsij_ss_A_002b} yields after straightforward by lengthy calculations
the asymptotic expansions for the invariants as $y^+\to0$ \tabrefsab{Tab_DTWT_s_DNSDepsij_ss_WA_001}
                                                                    {Tab_DTWT_s_DNSDepsij_ss_WA_002}.

%
%
%
%
%
%
%
%
%
\bibliographystyle{jfm}\footnotesize\bibliography{Aerodynamics,GV,GV_news}\normalsize
%
%
%
%
%
%
%
%
%

\end{document}